\documentclass[aps,onecolumn ]{revtex4}
\usepackage{amsfonts}
\usepackage{amsmath}
\usepackage{amssymb,epsf}
\usepackage{graphicx,color}

\begin{document}

\title{Phase transition of black holes in Brans--Dicke Born--Infeld gravity\\
through geometrical thermodynamics}
\author{S. H. Hendi$^{1,2}$\footnote{email address: hendi@shirazu.ac.ir}, M. S. Talezadeh$^{1}$ and Z.
Armanfard$^{1}$} \affiliation{$^1$ Physics Department and Biruni
Observatory, College of Sciences, Shiraz
University, Shiraz 71454, Iran\\
$^2$ Research Institute for Astronomy and Astrophysics of Maragha
(RIAAM), P. O. Box 55134-441, Maragha, Iran}

\begin{abstract}
Using the geometrical thermodynamic approach, we study phase
transition of Brans--Dicke Born--Infeld black holes. We apply
introduced methods and describe their shortcomings. We also use
the recently proposed new method and compare its results with
those of canonical ensemble. By considering the new method, we
find that its Ricci scalar diverges in the places of phase
transition and bound points. We also show that the bound point can
be distinguished from the phase transition points through the sign
of thermodynamical Ricci scalar around its divergencies.

\end{abstract}

\maketitle

\section{Introduction}

General relativity is accepted as a standard theory of gravitation
and is able to pass more observational tests \cite{GR}. Although,
this theory is successful in various domains, it cannot describe
some experimental evidences such as the accelerating expansion of
the Universe \cite{Expansion1,Expansion2,Expansion3}. Moreover,
the general relativity theory is not consistent with Mach's
principle nor Dirac's large number hypothesis
\cite{Mach&Dirac1,Mach&Dirac2}. In addition, one needs further
accurate observations to fully confirm (or disprove) the validity
of general relativity in the high curvature regime such as black
hole systems and other massive objects. Therefore, in recent
years, more attentions have been focused on alternative theories
of gravity. The most considerable alternative theories of gravity
is the scalar-tensor theories. One of the good examples of these
theories is Brans--Dicke (BD) theory which was introduced in
$1961$ to combine the Mach's principle with the Einstein's theory
of gravity \cite{BD}. It is worthwhile to mention that BD theory
is one of the modified theories of general relativity which can be
used for several cosmological problems like inflation, cosmic
acceleration and dark energy modeling
\cite{BD-Example1,BD-Example2,BD-Example3}. Also, it has a
customizable parameter ($\omega$) which indicates the strength of
coupling between the matter and scalar fields. The action of
$4-$dimensional BD theory can be written as
\begin{equation}
S=\frac{1}{16\pi }\int d^{4}x\sqrt{-g}\left(\Phi R-\frac{\omega }{\Phi}
(\nabla \Phi)^{2}\right),
\end{equation}
where $R$ and $\Phi$ are, respectively, the Ricci scalar and self
gravitating scalar field. It is interesting to note that $4-$dimensional
stationary vacuum BD solution is just the Kerr solution with a trivial
scalar field \cite{Gao4}. In addition, Cai and Myung proved that $4-$%
dimensional solution of BD-Maxwell theory reduces to the
Reissner--Nordstr\"{o}m solution with a constant scalar field
\cite{sh5a,sh5b,sh55a,sh55b}. However, the solutions of BD-Maxwell
gravity in higher dimensions will be reduced to the
Reissner--Nordstr\"{o}m solutions with a non-trivial scalar field
because of the fact that higher dimensional stress energy tensor
of Maxwell field is not traceless (conformally invariant). One of
the most prominent problems which makes BD theory
non-straightforward is the fact that the field equations of this
theory are highly nonlinear. To deal with this issue, one could
apply conformal transformation on known solutions of other
modified theories like dilaton gravity \cite{Dilaton}. For
instance, nonlinearly charged dilatonic black hole solutions and
their BD counterpart in an energy dependent
spacetime have been obtained by applying a conformal transformation \cite%
{BDvsDilaton}.

The first attempt for modifying the Maxwell theory to a consistent
theory for describing point charges was made in $1912$ by Gustav
Mie \cite{Mie1,Mie2}. After that, Born and Infeld introduced a
gauge-invariant nonlinear electrodynamic model to find a classical
theory of point-like charges with finite energy density \cite{BI}.
Born-Infeld (BI) theory was more interesting since it was obtained
by using loop correction analysis of Quantum Field theory.
Recently, Tseytlin has shown that BI theory can be derived as an
effective theory of some string theory models
\cite{Low-energy1,Low-energy2,Low-energy3,Low-energy4,Low-energy5,Low-energy6}.
Nowadays, the effects of BI electrodynamics coupled to various
gravity theories have been considered by many authors in the
context of black holes
\cite{Blackhole1,Blackhole2,Blackhole3,Blackhole4,Blackhole5,Blackhole6,
Blackhole7,Blackhole8,Blackhole9,Blackhole10,Blackhole11,Blackhole12,
Blackhole13,Blackhole14,Blackhole15,Blackhole16,Blackhole17,Blackhole18,
Blackhole19,Blackhole20,Blackhole21,Blackhole22,Blackhole23,Blackhole24,Blackhole25},
rotating black branes
\cite{Rotating1,Rotating2,Rotating3,Rotating4,Rotating5,Rotating6,Rotating7},
wormholes \cite{Wormhole1,Wormhole2,Wormhole3,Wormhole4},
superconductors \cite{Super1,Super2,Super3,Super4,Super5,Super6}
and other aspects of physics \cite{BIpmi1,BIpmi2}.

On the other side, black hole thermodynamics became an interesting
topic after the works of Hawking and Beckenstein
\cite{HP1,HP2,HP3,HP4,HP5,HP6}. Besides, based on the AdS/CFT
correspondence, black hole thermodynamics was considered as the
first step for constructing quantum gravity. In recent years,
phase transition and critical behavior of the black holes have
attracted more attentions among physicists. Generally, at the
critical point where phase transition occurs, one may find a
discontinuity of state space variable such as heat capacity
\cite{HC}. In addition to heat capacity, there are various
approaches for studying phase transition. One of such interesting
methods is based on geometrical technique. Geometrical
thermodynamic method was started by Gibbs and Caratheodory
\cite{Callen}. Regarding this method, one could build a phase
space by employing thermodynamical potential and its corresponding
extensive parameter. Meanwhile, divergence points of Ricci scalar
of thermodynamical metric provide information about phase
transition points of thermodynamical systems.

First time, Weinhold introduced a new metric on the equilibrium
thermodynamical phase space \cite{Weinhold1,Weinhold2} and after
that another thermodynamical metric was defined by Ruppeiner from
a different point of view \cite{Ruppeiner1,Ruppeiner2}. It is
worthwhile to mention that, there is a conformally relation
between Ruppeiner and Weinhold metrics with the inverse of
temperature as a conformal factor \cite{Salamon}. None of Weinhold
and Ruppeiner metrics were invariant under Legendre
transformation. Recently, Quevedo \cite{Quevedo1,Quevedo2} removed
some problems of Weinhold and Ruppeiner methods by proposing a
Legendre invariant thermodynamical metric. Although Quevedo could
solve some problems which previous metrics were involved with, it
has been confronted with another problems in some specific
systems. To solve these problems, a new method was proposed in
Ref. \cite{HPEM1,HPEM2,HPEM3} which is known as HPEM metric. It
was shown that HPEM metric is completely consistent with the
results of the heat capacity in canonical ensemble in different
gravitational systems.

In this paper, we are going to consider black hole solutions of BD-BI as
well as Einstein-BI-dilaton gravity and study their phase transition based
on geometrical thermodynamic methods. We compare our results with those of
other methods such as extended phase space thermodynamics.

%%%%%%%%%%%%%%%%%%%%%%%%%%%%%%%%%%%%%%%%%%%%%%%%%%%%%%%%%%%%%%%%%%%%%%%%%%%%%%%%%%%%%%%%%%%

\section{FIELD EQUATION AND CONFORMAL TRANSFORMATIONS \label{FE}}

The $(n+1)-$dimensional BD-BI theory action containing a scalar
field $\Phi
$ and a self-interacting potential $V(\Phi )$ is as follows%
\begin{equation}
I_{BD-BI}=-\frac{1}{16\pi }\int_{M}d^{n+1}x\sqrt{-g}\left( \Phi R-\frac{%
\omega }{\Phi }\left( \nabla \Phi \right) ^{2}-V\left( \Phi \right) +%
\mathcal{L}(\mathcal{F})\right),  \label{action}
\end{equation}
where $\omega $ is a coupling constant and $\mathcal{L}(\mathcal{F})$ is the
BI theory Lagrangian
\begin{equation}
\mathcal{L}(\mathcal{F})=4\beta ^{2}\left( 1-\sqrt{1+\frac{\mathcal{F}}{%
2\beta ^{2}}}\right),  \label{BI-Lagrangian}
\end{equation}
in which $\beta$ and $\mathcal{F}=F_{\mu \nu}F^{\mu \nu}$ are BI parameter
and Maxwell invariant, respectively. It is worth mentioning that $\mathcal{L}%
(\mathcal{F})$ will be reduced to the standard Maxwell form $\mathcal{L}(%
\mathcal{F})=-\mathcal{F}$ as $\beta \rightarrow \infty $. The field
equations of gravitational, scalar and electromagnetic fields can be
obtained by varying the action (\ref{action})
\begin{eqnarray}
G_{\mu \nu } &=&\frac{\omega }{\Phi ^{2}}\left( \nabla _{\mu }\Phi \nabla
_{\nu }\Phi -\frac{1}{2}g_{\mu \nu }(\nabla \Phi )^{2}\right) -\frac{V(\Phi )%
}{2\Phi }g_{\mu \nu }+\frac{1}{\Phi }\left( \nabla _{\mu }\nabla _{\nu }\Phi
-g_{\mu \nu }\nabla ^{2}\Phi \right)  \nonumber \\
&&+\frac{2}{\Phi }\left( \frac{F_{\mu \lambda }F_{\nu }^{\text{ }\lambda }}{%
\sqrt{1+\frac{\mathcal{F}}{2\beta ^{2}}}}+\frac{1}{4}g_{\mu \nu }\mathcal{L}(%
\mathcal{F})\right),  \label{FBD1} \\
\nabla ^{2}\Phi &=&\frac{1}{2\left[ \left( n-1\right) \omega +n\right] }%
\left( (n-1)\Phi \frac{dV(\Phi )}{d\Phi }-\left( n+1\right) V(\Phi )+\left(
n+1\right) \mathcal{L}(\mathcal{F})+\frac{4\mathcal{F}}{\sqrt{1+\frac{%
\mathcal{F}}{2\beta ^{2}}}}\right),  \label{FBD2}
\end{eqnarray}
\begin{equation}
\nabla _{\mu }\left( \frac{F^{\mu \nu }}{\sqrt{1+\frac{\mathcal{F}}{2\beta
^{2}}}}\right) =0.  \label{FBD3}
\end{equation}

It is not easy to solve Eqs. (\ref{FBD1})-(\ref{FBD3}) because
there exist second order of scalar field in the denominator of
field equation (\ref{FBD1}). In order to overcome such a problem,
we can use a suitable conformal transformation and convert the
BD-BI theory to the Einstein-BI-dilaton gravity. The suitable
conformal transformation is as follows
\begin{equation}
\bar{g}_{\mu \nu }=\Phi ^{2/(n-1)}g_{\mu \nu },  \label{CT}
\end{equation}
\begin{eqnarray}
\bar{\Phi} &=&\frac{n-3}{4\alpha }\ln \Phi ,  \label{Phibar} \\
\alpha &=&(n-3)/\sqrt{4(n-1)\omega +4n}.  \label{alpha}
\end{eqnarray}

The Einstein-BI-dilaton gravity action and its related field equations can
be obtained from the BD-BI action and its related field equations by
applying the mentioned conformal transformation \cite{BDvsDilaton}
\begin{equation}
\overline{I}_{G}=-\frac{1}{16\pi }\int_{\mathcal{M}}d^{n+1}x\sqrt{-\overline{%
g}}\left\{ \overline{\mathcal{R}}-\frac{4}{n-1}(\overline{\nabla }\overline{%
\Phi })^{2}-\overline{V}(\overline{\Phi })+\overline{L}\left( \overline{%
\mathcal{F}},\overline{\Phi }\right) \right\} ,  \label{con-ac}
\end{equation}%
\begin{eqnarray}
\overline{\mathcal{R}}_{\mu \nu } &=&\frac{4}{n-1}\left( \overline{\nabla }%
_{\mu }\overline{\Phi }\overline{\nabla }_{\nu }\overline{\Phi }+\frac{1}{4}%
\overline{V}(\overline{\Phi })\overline{g}_{\mu \nu }\right) -\frac{1}{n-1}%
\overline{L}(\overline{\mathcal{F}},\overline{\Phi })\overline{g}_{\mu \nu }+%
\frac{2e^{-\frac{4\alpha \overline{\Phi }}{n-1}}}{\sqrt{1+\overline{Y}}}%
\left( \overline{F}_{\mu \eta }\overline{F}_{\nu }^{\eta }-\frac{\overline{%
\mathcal{F}}}{n-1}\overline{g}_{\mu \nu }\right) ,  \label{FE1} \\
\overline{\nabla }^{2}\overline{\Phi } &=&\frac{n-1}{8}\frac{\partial
\overline{V}(\overline{\Phi })}{\partial \overline{\Phi }}+\frac{\alpha }{%
2(n-3)}\left( (n+1)\overline{L}(\overline{\mathcal{F}},\overline{\Phi })+%
\frac{4e^{-\frac{4\alpha \overline{\Phi }}{n-1}}\overline{\mathcal{F}}}{%
\sqrt{1+\overline{Y}}}\right) ,  \label{FE2}
\end{eqnarray}%
\begin{equation}
\overline{\nabla }_{\mu }\left( \frac{e^{-\frac{4\alpha \overline{\Phi }}{n-1%
}}}{\sqrt{1+\overline{Y}}}\overline{F}^{\mu \nu }\right) =0,  \label{FE3}
\end{equation}%
where $\overline{\nabla }$ is the covariant differentiation with respect to
the metric $\overline{g}_{\mu \nu }$and $\overline{\mathcal{R}}$ is its
Ricci scalar. The potential $\overline{V}\left( \overline{\Phi }\right) $
and the Lagrangian $\overline{L}\left( \overline{F},\overline{\Phi }\right) $
will take the following forms \cite{BDvsDilaton}
\begin{equation}
\overline{V}(\overline{\Phi })=\Phi ^{-(n+1)/(n-1)}V(\Phi ),  \label{poten}
\end{equation}%
\begin{equation}
\overline{L}\left( \overline{\mathcal{F}},\overline{\Phi }\right) =4\beta
^{2}e^{-4\alpha \left( n+1\right) \overline{\Phi }/\left[ \left( n-1\right)
\left( n-3\right) \right] }\left( 1-\sqrt{1+\frac{e^{16\alpha \overline{\Phi
}/\left[ \left( n-1\right) \left( n-3\right) \right] }\overline{\mathcal{F}}%
}{2\beta ^{2}}}\right) .  \label{LFP}
\end{equation}

In the limits of $\beta \rightarrow \infty $\ and $\beta \rightarrow 0$, the
Lagrangian will be $\overline{L}\left( \overline{\mathcal{F}},\overline{\Phi
}\right) =-e^{-4\alpha \overline{\Phi }/\left( n-1\right) }\overline{%
\mathcal{F}}\ $and$\ \overline{L}\left( \overline{\mathcal{F}},\overline{%
\Phi }\right) \rightarrow 0$, respectively, as expected. In
previous equations, we have used the following notations
\begin{equation}
\overline{L}\left( \overline{\mathcal{F}},\overline{\Phi }\right) =4\beta
^{2}e^{-4\alpha \left( n+1\right) \overline{\Phi }/\left[ \left( n-1\right)
\left( n-3\right) \right] }\overline{L}\overline{(Y}),  \label{LFP2}
\end{equation}%
\begin{eqnarray}
\overline{L}\left( \overline{Y}\right) &=&1-\sqrt{1+\overline{Y}},
\label{L(Y)} \\
\overline{Y} &=&\frac{e^{16\alpha \overline{\Phi }/\left[ \left( n-1\right)
\left( n-3\right) \right] }\overline{\mathcal{F}}}{2\beta ^{2}}.  \label{Y}
\end{eqnarray}

By considering the conformal relation between these two theories it can be
understood that if \ $\left( \overline{g}_{\mu \nu },\overline{F}_{\mu \nu },%
\overline{\Phi }\right) $ \ are the solutions to the field equations of
Einstein-BI-dilaton gravity (\ref{FE1})-(\ref{FE3}), then the solutions of
BD-BI theory could be obtained by the following form
\begin{equation}
\left[ g_{\mu \nu },F_{\mu \nu },\Phi \right] =\left[ \exp \left( -\frac{%
8\alpha \overline{\Phi }}{\left( n-1\right) (n-3)}\right) \overline{g}_{\mu
\nu },\overline{F}_{\mu \nu },\exp \left( \frac{4\alpha \overline{\Phi }}{n-3%
}\right) \right] .  \label{BDsol}
\end{equation}

\subsection{Black hole solutions in Einstein-BI-dilaton gravity and BD-BI
theory \label{Sol}}

\subsubsection{\textbf{Einstein frame:}}

In this section, we briefly obtain the Einstein-BI-dilaton gravity
solutions and then by using the conformal transformation, we
calculate the solutions of BD-BI theory \cite{HMAT}. We assume the
following metric with various horizon topology
\begin{equation}
d\overline{s}^{2}=-Z(r)dt^{2}+\frac{dr^{2}}{Z(r)}+r^{2}R^{2}(r)d\Omega
_{k}^{2},  \label{metric}
\end{equation}
where $d\Omega_{k}^{2}$ is an $(n-1)$-dimensional hypersurface of Euclidean
metric with constant curvature $(n-1)(n-2)k$ and volume $\varpi_{n-1}$ with
the following explicit form
\begin{equation}
d\Omega _{k}^{2}=\left\{
\begin{array}{cc}
d\theta _{1}^{2}+\sum\limits_{i=2}^{n-1}\prod\limits_{j=1}^{i-1}\sin
^{2}\theta _{j}d\theta _{i}^{2} & k=1 \\
d\theta _{1}^{2}+\sinh ^{2}\theta _{1}d\theta _{2}^{2}+\sinh ^{2}\theta
_{1}\sum\limits_{i=3}^{n-1}\prod\limits_{j=2}^{i-1}\sin ^{2}\theta
_{j}d\theta _{i}^{2} & k=-1 \\
\sum\limits_{i=1}^{n-1}d\phi _{i}^{2} & k=0%
\end{array}
\right. .  \label{k}
\end{equation}

In order to obtain consistent solutions, we should consider a
suitable functional form for the potential,
$\mathbf{\overline{V}}(\overline{\Phi})$. It was shown that the
proper potential is a Liouville-type one with both topological and
BI correction terms, as \cite{BDvsDilaton}
\begin{equation}
\mathbf{\overline{V}}(\overline{\Phi })=2\Lambda \exp \left( \frac{4\alpha
\overline{\Phi }}{n-1}\right) +\frac{k(n-1)(n-2)\alpha ^{2}}{b^{2}\left(
\alpha ^{2}-1\right) }\exp \left( \frac{4\overline{\Phi }}{(n-1)\alpha }%
\right) +\frac{W(r)}{\beta ^{2}}.  \label{liovilpoten}
\end{equation}

It is notable to mention that in the limit of $\alpha \rightarrow 0$
(absence of dilaton field) and $\beta \rightarrow \infty $, $\mathbf{%
\overline{V}}(\overline{\Phi })$ reduces to $2\Lambda$, as expected \cite%
{pakravan}. Now, regarding the field equations (\ref{FE1})-(\ref{FE3}),
metric (\ref{metric}) and the potential $\mathbf{\overline{V}}(\overline{\Phi%
})$, it is a matter of calculation to show that
\begin{eqnarray}
F_{tr} &=&E(r)=\frac{qe^{\left( \frac{4\alpha \overline{\Phi }(r)}{n-1}%
\right) }}{[r R(r)]^{(n-1)}\sqrt{1+\frac{e^{(\frac{8\alpha \overline{\Phi }%
(r)}{n-3})}q^{2}[r R(r)]^{-2(n-1)}}{\beta ^{2}}}},  \label{E} \\
\overline{\Phi } &=&\frac{(n-1)\alpha }{2(1+\alpha ^{2})}\ln \left( \frac{b}{%
r}\right),  \label{phi}
\end{eqnarray}%
\begin{equation}
W(r)=\frac{4q(n-1)\beta ^{2}R(r)}{\left( 1+\alpha ^{2}\right) r^{\gamma
}b^{n\gamma }}\int \frac{E(r)}{r^{n(1-\gamma )-\gamma }}dr+\frac{4\beta ^{4}%
}{R(r)^{\frac{2(n+1)}{n-3}}}\left( 1-\frac{E(r)R(r)^{(n-3)}}{qr^{1-n}}%
\right) -\frac{4q\beta ^{2}E(r)}{r^{n-1}}(\frac{r}{b})^{\gamma (n-1)},
\label{W}
\end{equation}%
\begin{eqnarray}
Z(r) &=&-\frac{k\left( n-2\right) \left( \alpha ^{2}+1\right)
^{2}\left( \frac{r}{b}\right)^{2\gamma }}{\left( \alpha
^{2}+n-2\right) \left( \alpha
^{2}-1\right) }+\left( \frac{(1+\alpha ^{2})^{2}r^{2}}{(n-1)}\right) \frac{%
2\Lambda \left( \frac{r}{b}\right) ^{-2\gamma }}{(\alpha ^{2}-n)}-\frac{m}{%
r^{(n-1)(1-\gamma )-1}}  \nonumber \\
&&-\frac{4(1+\alpha ^{2})^{2}q^{2}(\frac{r}{b})^{2\gamma (n-2)}}{(n-\alpha
^{2})r^{2(n-2)}}\left( \frac{1}{2(n-1)}\digamma _{1}(\eta )-\frac{1}{\alpha
^{2}+n-2}\digamma _{2}(\eta )\right),  \label{f}
\end{eqnarray}
\begin{equation}
R(r)= \exp \left( \frac{2\alpha \overline{\Phi }}{n-1}\right) =\left( \frac{b%
}{r}\right)^{\gamma },  \label{R(r)}
\end{equation}
where $m$ is an integration constant related to mass and $b$ is another
constant related to scalar field, and
\begin{eqnarray}
\digamma _{1}(\eta ) &=&\text{ }_{2}F_{1}\left( \left[ \frac{1}{2},\frac{%
(n-3)\Upsilon }{\alpha ^{2}+n-2}\right] ,\left[ 1+\frac{(n-3)\Upsilon }{%
\alpha ^{2}+n-2}\right] ,-\eta \right) ,  \nonumber \\
\digamma _{2}(\eta ) &=&\text{ }_{2}F_{1}\left( \left[ \frac{1}{2},\frac{%
(n-3)\Upsilon }{2(n-1)}\right] ,\left[ 1+\frac{(n-3)\Upsilon }{2(n-1)}\right]
,-\eta \right) ,  \nonumber \\
\Upsilon &=&\frac{\alpha ^{2}+n-2}{2\alpha ^{2}+n-3},  \nonumber \\
\eta &=&\frac{q^{2}(\frac{r}{b})^{2\gamma (n-1)(n-5)/(n-3)}}{\beta
^{2}r^{2(n-1)}},  \nonumber \\
\gamma&=& \frac{\alpha^2}{1+\alpha^2},   \nonumber
\end{eqnarray}

It is worthwhile to mention that, the dilatonic Maxwell solutions \cite%
{Sheikhi} can be achieved from the obtained solutions in the limit
of $\beta \rightarrow \infty $. The divergencies of scalar
curvatures at the origin guarantee the existence of singularity.
We interpret such a singularity as black hole since it is covered
by an event horizon \cite{BDvsDilaton}.

\subsubsection{\textbf{Jordan frame:}}

To obtain the black hole solutions of BD-BI theory, first, by
using the conformal transformation (\ref{poten}) the
$\mathbf{V}(\Phi )$ would be
\begin{equation}
\mathbf{V}(\Phi )=2\Lambda \Phi ^{2}+\frac{k(n-1)(n-2)\alpha ^{2}}{%
b^{2}\left( \alpha ^{2}-1\right) }\Phi ^{\lbrack (n+1)(1+\alpha
^{2})-4]/[(n-1)\alpha ^{2}]}+\Phi ^{(n+1)/(n-1)}\frac{W(r)}{\beta ^{2}}.
\label{V(phi)}
\end{equation}

Also, by considering the following $(n+1)$-dimensional metric
\begin{equation}
ds^{2}=-A(r)dt^{2}+\frac{dr^{2}}{B(r)}+r^{2}H^{2}(r)d\Omega _{k}^{2},
\label{metric1}
\end{equation}
one can find the following solutions through conformal transformation
\begin{eqnarray}
A(r) &=&\left( \frac{r}{b}\right) ^{4\gamma /\left( n-3\right) }Z\left(
r\right) ,  \label{A(r)} \\
B(r) &=&\left( \frac{r}{b}\right) ^{-4\gamma /\left( n-3\right) }Z\left(
r\right) ,  \label{B(r)} \\
H(r) &=&\left( \frac{r}{b}\right) ^{-\gamma (\frac{n-5}{n-3})},  \label{H(r)}
\\
\Phi \left( r\right) &=&\left( \frac{r}{b}\right) ^{-\frac{2\gamma \left(
n-1\right) }{n-3}}.  \label{Phi}
\end{eqnarray}

It is notable that, like Einstein frame, these solutions can be
interpreted as black holes which are covered by event horizon.

\section{Thermodynamic properties: Dilatonic-BI vs BD-BI black holes \label{p-vb}}

\subsection{Thermodynamic quantities:}

In the following, we give a brief review regarding thermodynamic quantities
of the black hole solutions in both frames. The Hawking temperature of the
black hole can be obtained by using the surface gravity interpretation ($%
\kappa$) through the following relation
\begin{equation}
T=\frac{\kappa }{2\pi }=\frac{1}{2\pi }\sqrt{-\frac{1}{2}\left( \nabla _{\mu
}\chi _{\nu }\right) \left( \nabla ^{\mu }\chi ^{\nu }\right) }=\left\{
\begin{array}{cc}
\frac{Z^{\prime }(r_{+})}{4\pi }, & \text{dilatonic BI} \\
\nonumber &  \\
\frac{1}{4\pi }\sqrt{\frac{B(r)}{A(r)}}A^{\prime }(r_{+}), & \text{BD-BI}%
\end{array}%
\right. ,  \label{Td1}
\end{equation}%
in which $\chi =\partial /\partial t$ is the time like null Killing vector.
It is easy to show that the Hawking temperature in both frames is uniform as
\begin{equation}
T=\frac{\left( \alpha ^{2}+1\right) }{2\pi \left( n-1\right) }\left[ \frac{%
-k\left( n-2\right) (n-1)}{2\left( \alpha ^{2}-1\right) r_{+}}\left( \frac{b%
}{r_{+}}\right) ^{-2\gamma }-\Lambda r_{+}\left(
\frac{b}{r_{+}}\right) ^{2\gamma }+\Gamma_{+} \right] ,\text{
dilatonic BI \& BD-BI}  \label{Td2}
\end{equation}%
where
\begin{equation}
\Gamma_{+} =-\frac{\left( \alpha ^{2}+1\right) ^{2}q^{2}}{2\pi
(n-1)}\left( \frac{r_{+}}{b}\right) ^{2\gamma \left( n-2\right)
}r_{+}^{3-2n}\digamma _{1}(\eta_{+} ).  \label{GAMMA}
\end{equation}

Following Refs. \cite{BDvsDilaton,Cai}, the finite mass and
entropy of the black hole in both Einstein and Jordan frames are
\begin{eqnarray}
M &=&\frac{\varpi _{n-1}b^{(n-1)\gamma }}{16\pi }\left(
\frac{n-1}{1+\alpha
^{2}}\right) m,  \label{Md} \\
S &=&\frac{\varpi _{n-1}b^{(n-1)\gamma }}{4}r_{+}^{(n-1)\left(
1-\gamma \right)}.  \label{Sd}
\end{eqnarray}

In addition, the electric charge $Q$ of the black holes can be obtained via
the Gauss's law
\begin{equation}
Q=\frac{q}{4\pi }.  \label{Qd}
\end{equation}

Also, one can obtain the electric potential as
\begin{eqnarray}
U &=&\left(\frac{r_{+}}{b}\right)^{4\gamma+1 }\frac{b\beta (\alpha
^{2}+1)}{ (5\alpha ^{2}+1)}\;{}_{2}F_{1}\left( \left[
\frac{1}{2},\frac{5\alpha ^{2}+1}{6(2\alpha ^{2}+1)}\right]
,\left[ \frac{17\alpha ^{2}+7}{6(2\alpha ^{2}+1)}\right]
,\frac{\beta
^{2}b^{6}}{q^{2}}\left(\frac{r_{+}}{b}\right)^{6\gamma+6 }\right)
.\label{U}
\end{eqnarray}

It is straightforward to show that the mentioned conserved and
thermodynamical quantities satisfy the first law of thermodynamics as
\begin{equation}
dM=TdS+UdQ.  \label{1stLAW}
\end{equation}

\subsection{Heat capacity and thermal stability}

Here, we want to investigate thermal stability of the black holes. Due to
the set of state functions and thermodynamic variables of a system, one may
study the thermodynamic stability from different points of view through
various ensembles. One of the common methods to study phase transition is
regarding the canonical ensemble. In this ensemble, thermal stability of a
system will be ensured by positivity of the heat capacity. One can obtain
the heat capacity relation with fixed charge as
\begin{equation}
C_{Q}=\frac{\left( \frac{\partial M}{\partial S}\right) _{Q}}{\left( \frac{%
\partial ^{2}M}{\partial S^{2}}\right) _{Q}}=\frac{M_{S}}{M_{SS}}=T\left(
\frac{\partial S}{\partial T}\right) _{Q},  \label{CQ}
\end{equation}%
where $M_{S}=\frac{\partial M}{\partial S}$ and $M_{SS}=\frac{\partial ^{2}M%
}{\partial S^{2}}$.

From the nominator of heat capacity, it is evident that the
temperature ($M_{S}$) has crucial role on the sign of $C_{Q}$. In
addition, divergence points of heat capacity are indicating second
order phase transition. Hence, these divergencies are utilized for
calculating critical values and investigating the critical
behavior of the black hole. Now, for studying phase transition, we
introduce various geometrical thermodynamic methods and compare
their results with those of arisen from the heat capacity.

\subsection{GEOMETRICAL STUDY OF THE PHASE TRANSITION}

One of the basic motivations for considering the geometrical
thermodynamics comes from the fact that this formalism helps us to
describe in an invariant way the thermodynamic properties of a
given thermodynamical system in terms of geometric structures.
Also, this method is a strong machinery for describing phase
transition of the black holes. Another motivation is to give an
independent picture regarding thermodynamical aspects of a system.
In addition to some useful information about bound points, phase
transitions and thermal stability conditions, this method contains
information regarding molecular interaction around phase
transitions for thermodynamical systems. In other words, by
studying the sign of thermodynamical Ricci scalar around phase
transition points, one can extract information whether interaction
is repulsive or attractive. Based on such motivations, it will be
interesting to investigate black hole phase transition in the
context of geometrical thermodynamics, as an independent approach.

In order to study the phase transition, one can employ
thermodynamical quantities to build geometrical spacetime. There
are several metrics in the context of geometrical thermodynamics
which one can use them to study phase transition and critical
behavior. The well-known thermodynamical metrics are Weinhold,
Ruppeiner, Quevedo and HPEM as the recently proposed method. As we
mentioned, in some specific types of systems the Weinhold,
Ruppeiner and Quevedo metrics are not applicable and they will
face some problems. Here, we want to discuss these thermodynamical
metrics and their possible mismatched problems.

\begin{figure}[tbp]
$%
\begin{array}{cc}
\epsfxsize=5.5cm \epsffile{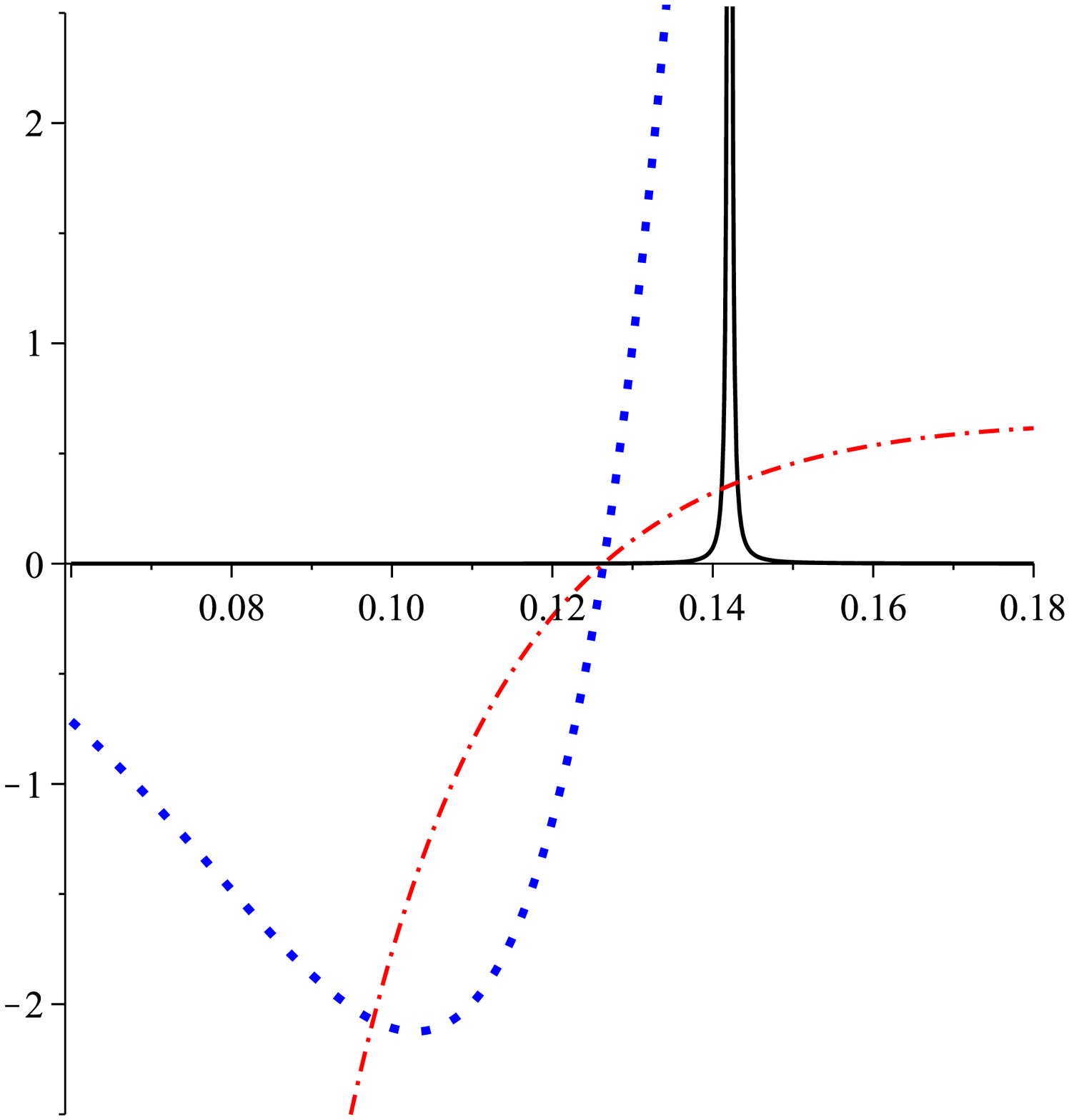} & \epsfxsize=5.5cm \epsffile{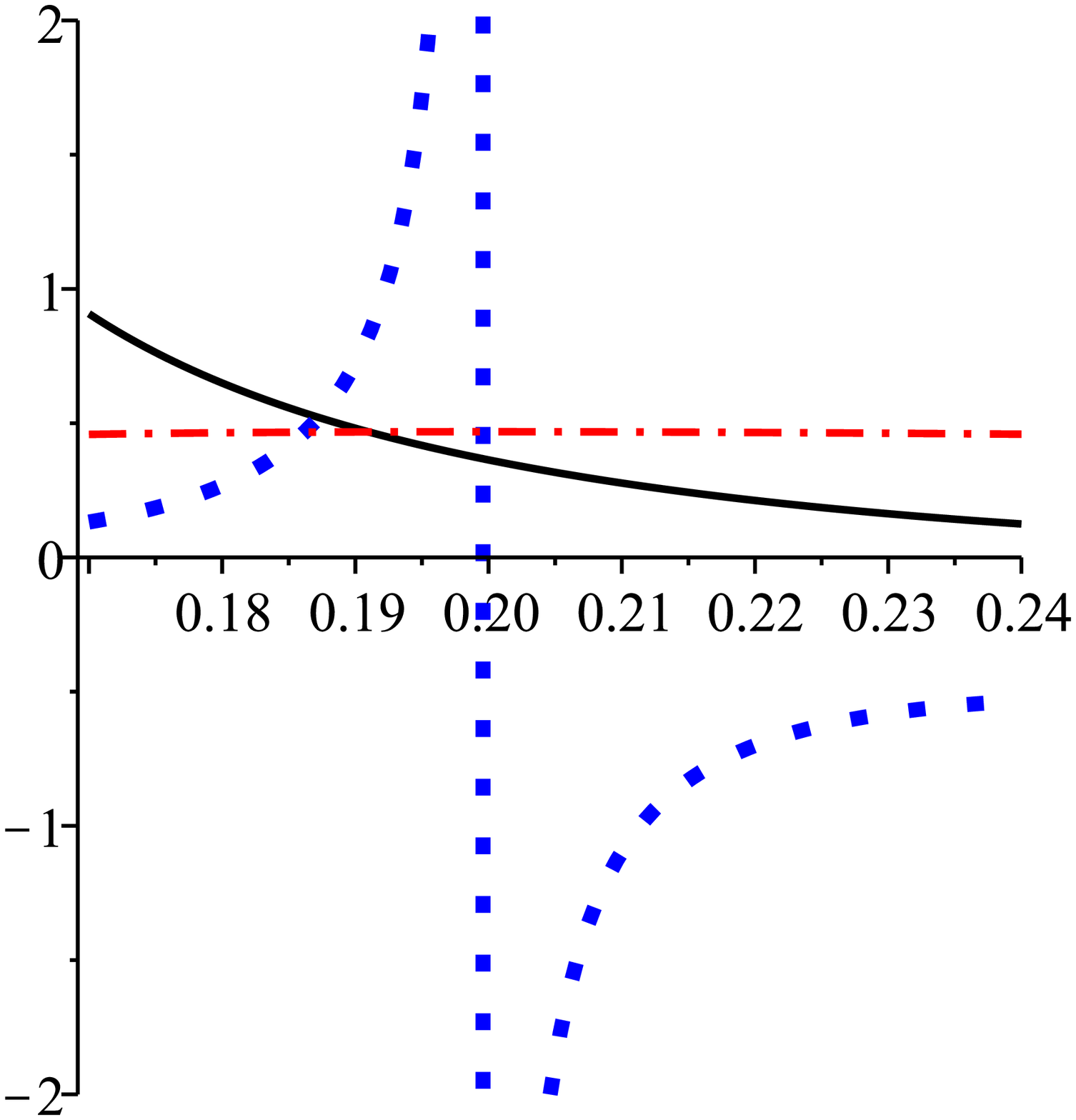} %
\epsfxsize=5.5cm \epsffile{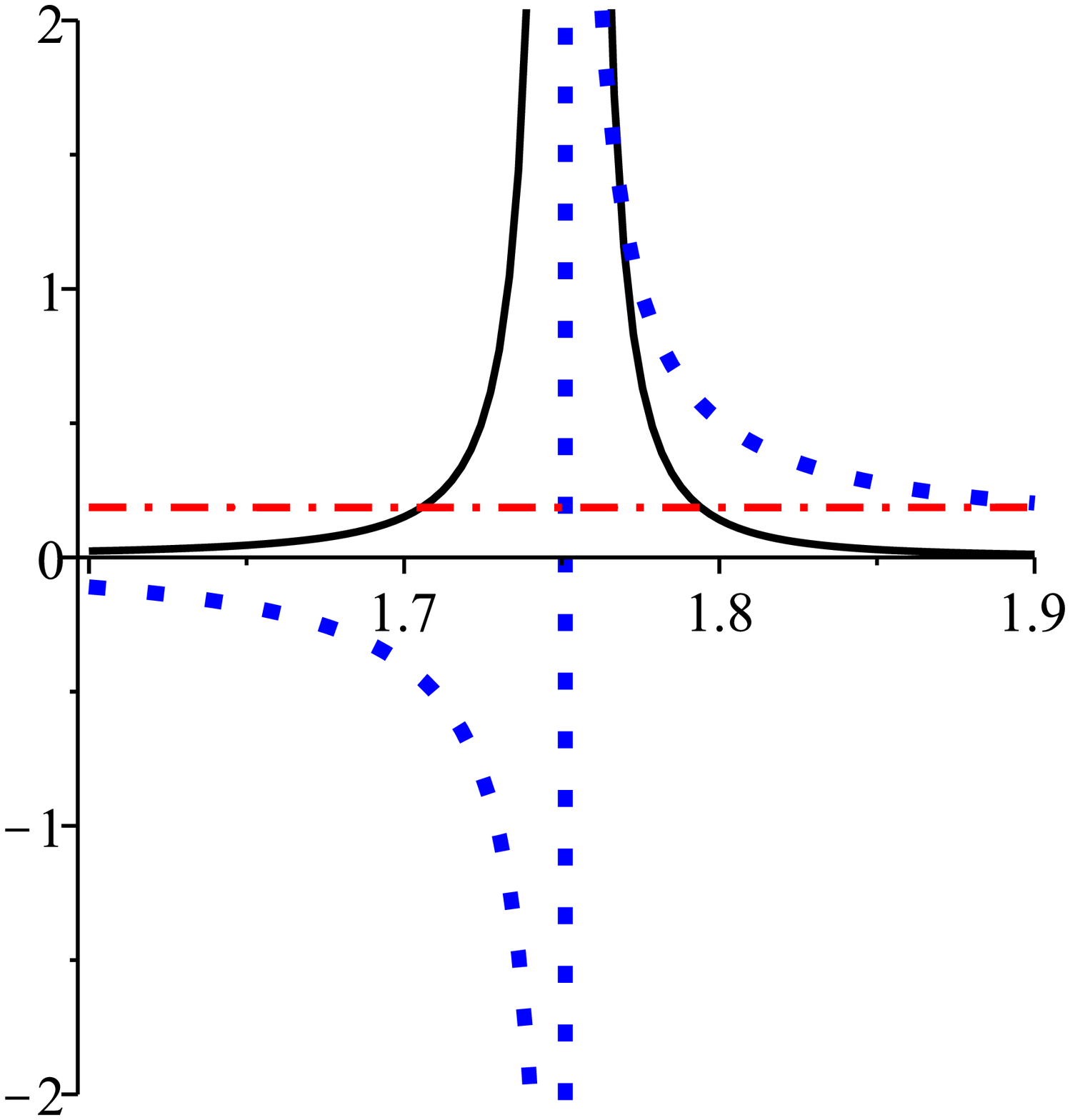}%
\end{array}
$%
\caption{\textbf{Weinhold metric:} $\mathcal{R}$ (continuous line), $C_{Q}$
(dotted line) and {T} (dot dashed) versus $r_{+}$ for $\Lambda =-1$, $n=4$, $%
q=0.1$, $b=1$, $\protect\omega =10$ and $\protect\beta =1.5$.
\emph{"Note: All three panels are plotted with the same
parameters, but different regions and scales."}} \label{FigW}
\end{figure}

Thermodynamical metric was first introduced by Weinhold
\cite{Weinhold1,Weinhold2}. This thermodynamical metric is given
by
\begin{equation}
dS_{W}^{2}=g_{ab}^{W}dX^{a}dX^{b},  \label{Weinmetric}
\end{equation}
where $g_{ab}^{W}=\partial ^{2}M\left( X^{c}\right) /\partial
X^{a}\partial X^{b}$, $X^{a}\equiv X^{a}\left( S,N^{i}\right) $
and $N^{i}$ denotes other extensive variables of the system. By
calculating $M$ as a function of extensive quantities (such as
entropy and electric charge) and using Weinhold metric
(\ref{Weinmetric}), one can find the Ricci scalar. It is expected
that the singular points of the Weinhold Ricci scalar match to the
root or divergence points of the heat capacity, to indicate the
bound point or the phase transition ones. We plot Fig. \ref{FigW}
to investigate the mentioned behavior.

After that, Ruppeiner \cite{Ruppeiner1,Ruppeiner2} has defined
another thermodynamical metric with the following form
\begin{equation}
dS_{R}^{2}=g_{ab}^{R}dX^{a}dX^{b},  \label{Rupp}
\end{equation}
where $g_{ab}^{R}=-\partial ^{2}S\left( X^{c}\right) /\partial X^{a}\partial
X^{b}$ and $X^{a}\equiv X^{a}\left( M,N^{i}\right) $.

\begin{figure}[tbp]
$%
\begin{array}{cc}
\epsfxsize=7cm \epsffile{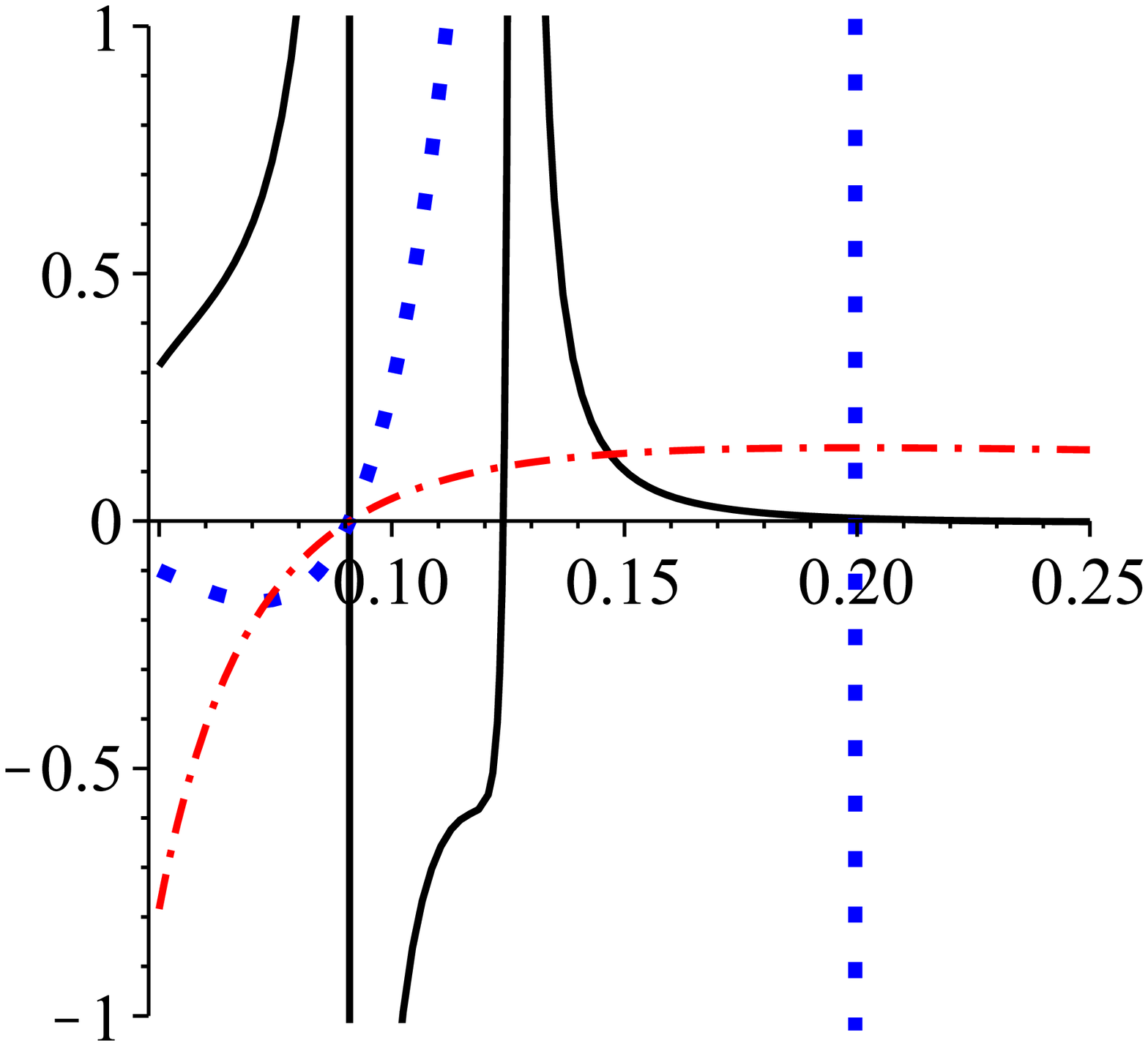} & \epsfxsize=7cm \epsffile{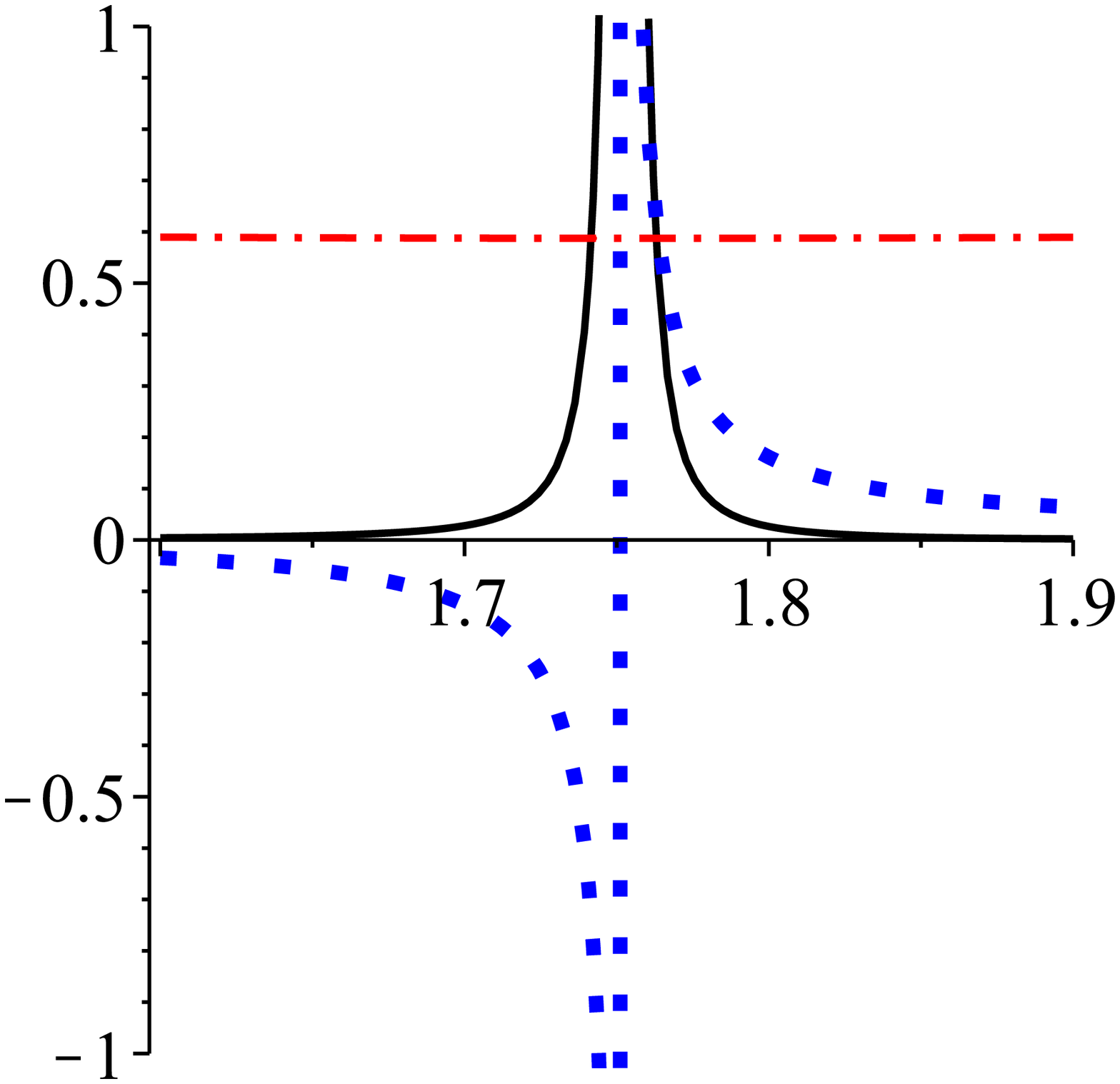}%
\end{array}
$%
\caption{\textbf{Ruppeiner metric:} $\mathcal{R}$ (continuous line), $C_{Q}$
(dotted line) and {T} (dot dashed) versus $r_{+}$ for $\Lambda =-1$, $n=4$, $%
q=0.1$, $b=1$, $\protect\omega =10$ and $\protect\beta =1.5$.
\emph{"Note: Both panels are plotted with the same parameters, but
different regions and scales."}} \label{FigR}
\end{figure}

In the Ruppeiner metric, thermodynamical potential is entropy. It
is worthwhile mentioning that these two metrics are conformally
related to each other \cite{Salamon}. We plot Fig. \ref{FigR} to
show that the Ruppeiner Ricci scalar divergencies are not matched
with those of heat capacity.

As we have shown, calculating thermodynamical Ricci scalar of
these two thermodynamical metrics indicates that the results were
not completely consistent with the results of heat capacity in the
canonical ensemble. In order to remove some failures of the
Weinhold and Ruppeiner metrics, recently, another metric which is
Legendre invariant has been introduced by Quevedo
\cite{Quevedo1,Quevedo2}. The Quevedo metric has the following
form
\begin{equation}
ds_{Q}^{2}=\Omega (-M_{SS}dS^{2}+M_{QQ}dQ^{2})  \label{quevedo}
\end{equation}
where the conformal coefficient $\Omega$ is
\begin{equation}
\Omega =\left( SM_{S}+QM_{Q}\right) .  \label{Omega}
\end{equation}

\begin{figure}[tbp]
$%
\begin{array}{cc}
\epsfxsize=7cm \epsffile{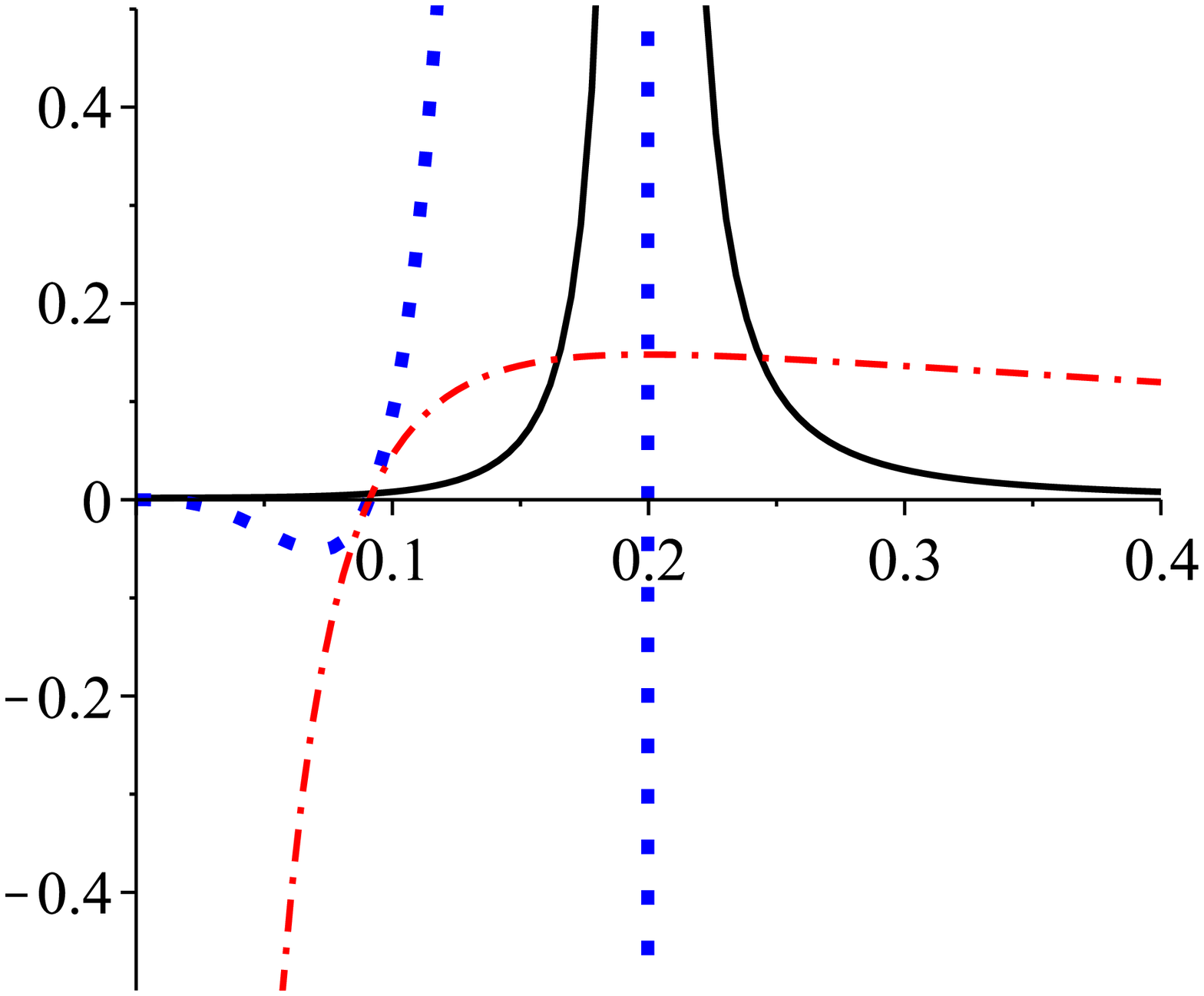} & \epsfxsize=7cm
\epsffile{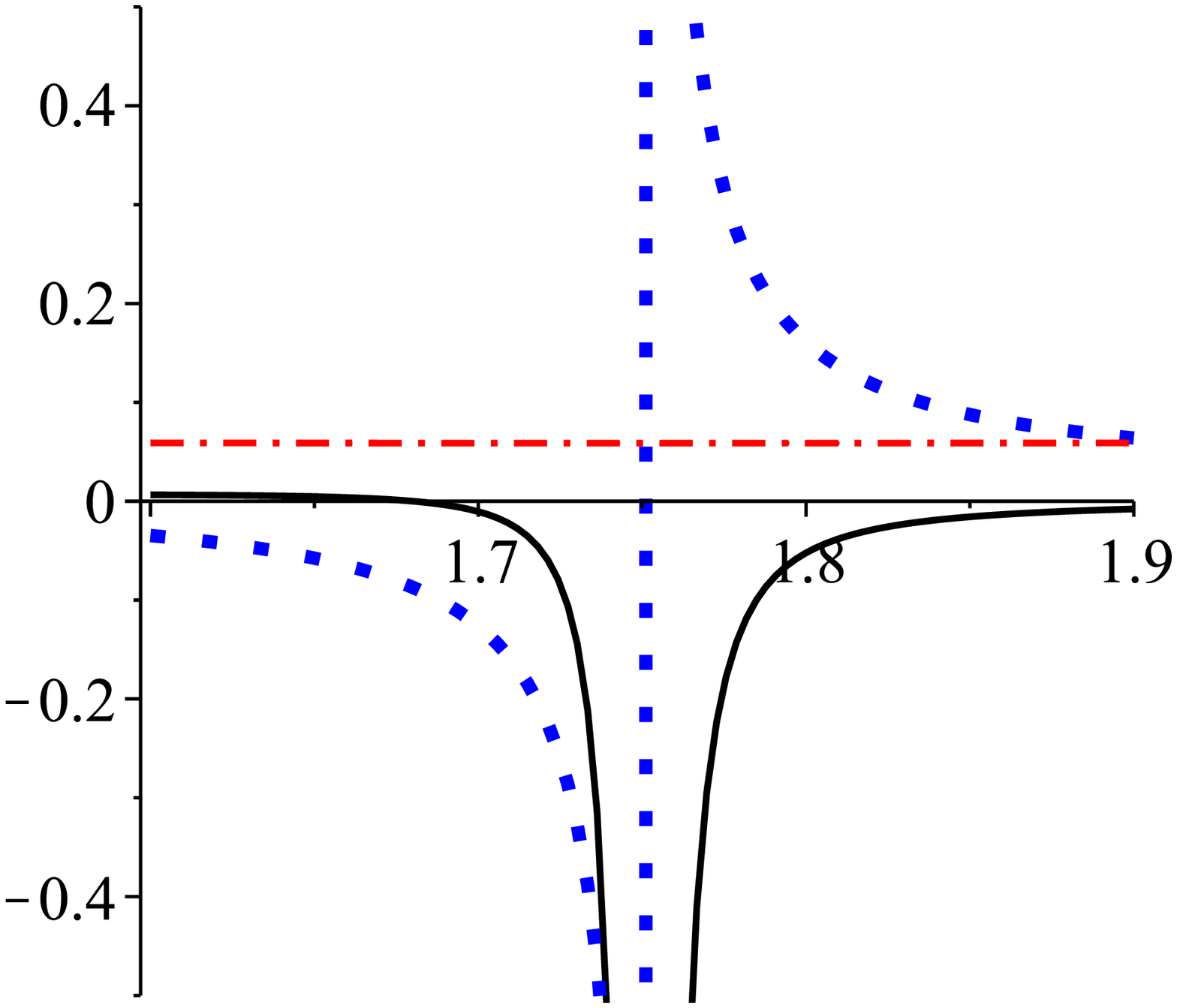}
\end{array}
$%
\caption{\textbf{Quevedo metric:} $\mathcal{R}$ (continuous line), $C_{Q}$
(dotted line) and {T} (dot dashed) versus $r_{+}$ for $\Lambda =-1$, $n=4$, $%
q=0.1$, $b=1$, $\protect\omega =10$ and $\protect\beta =1.5$.
\emph{"Note: Both panels are plotted with the same parameters, but
different regions."}} \label{FigQ}
\end{figure}

Considering Figs. \ref{FigW}--\ref{FigQ}, we find that by using
these three well-known metrics, there is at least a mismatch
between heat capacity divergencies and thermodynamical Ricci
scalar divergencies (of these three metrics). Therefore, these
metrics are not appropriate tools for investigation of our black
hole phase transitions and related critical behavior. In other
words, the method of geometrical thermodynamics which has been
reported in \cite{Niu} is not an applicable method in the scalar
field theory.

\begin{figure}[tbp]
$%
\begin{array}{cc}
\epsfxsize=6cm \epsffile{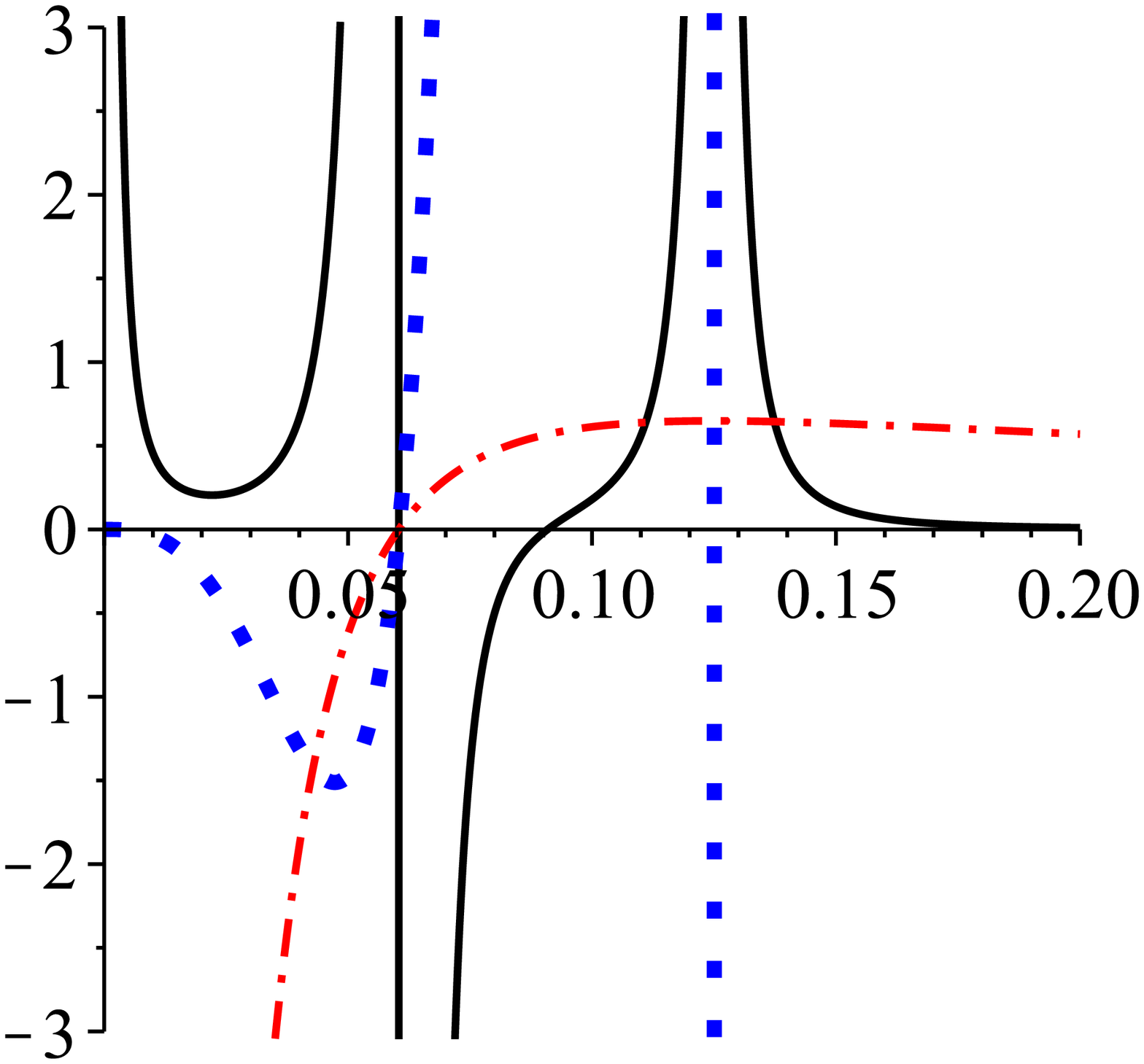} & \epsfxsize=6cm %
\epsffile{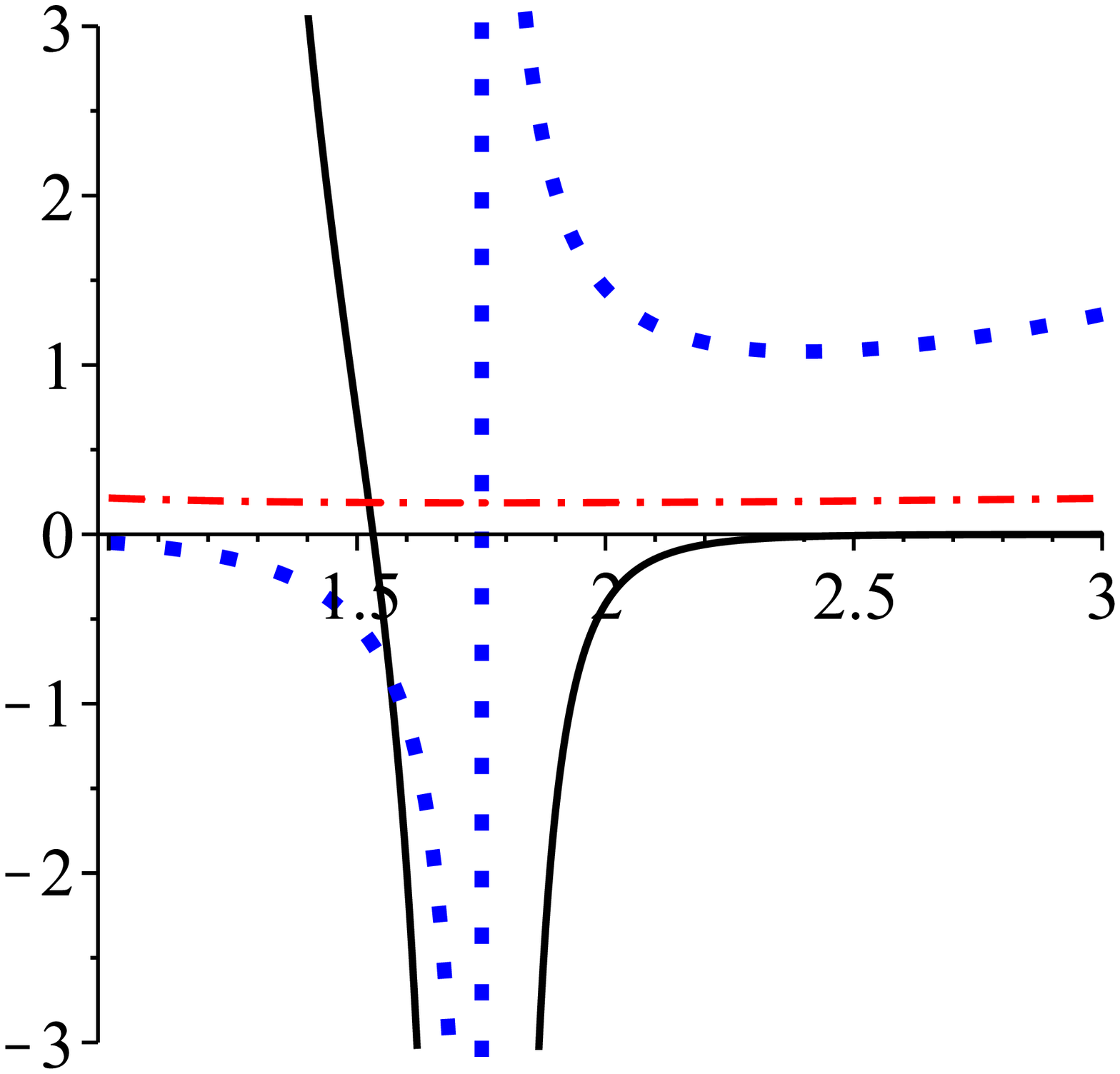} \\
\epsfxsize=6cm \epsffile{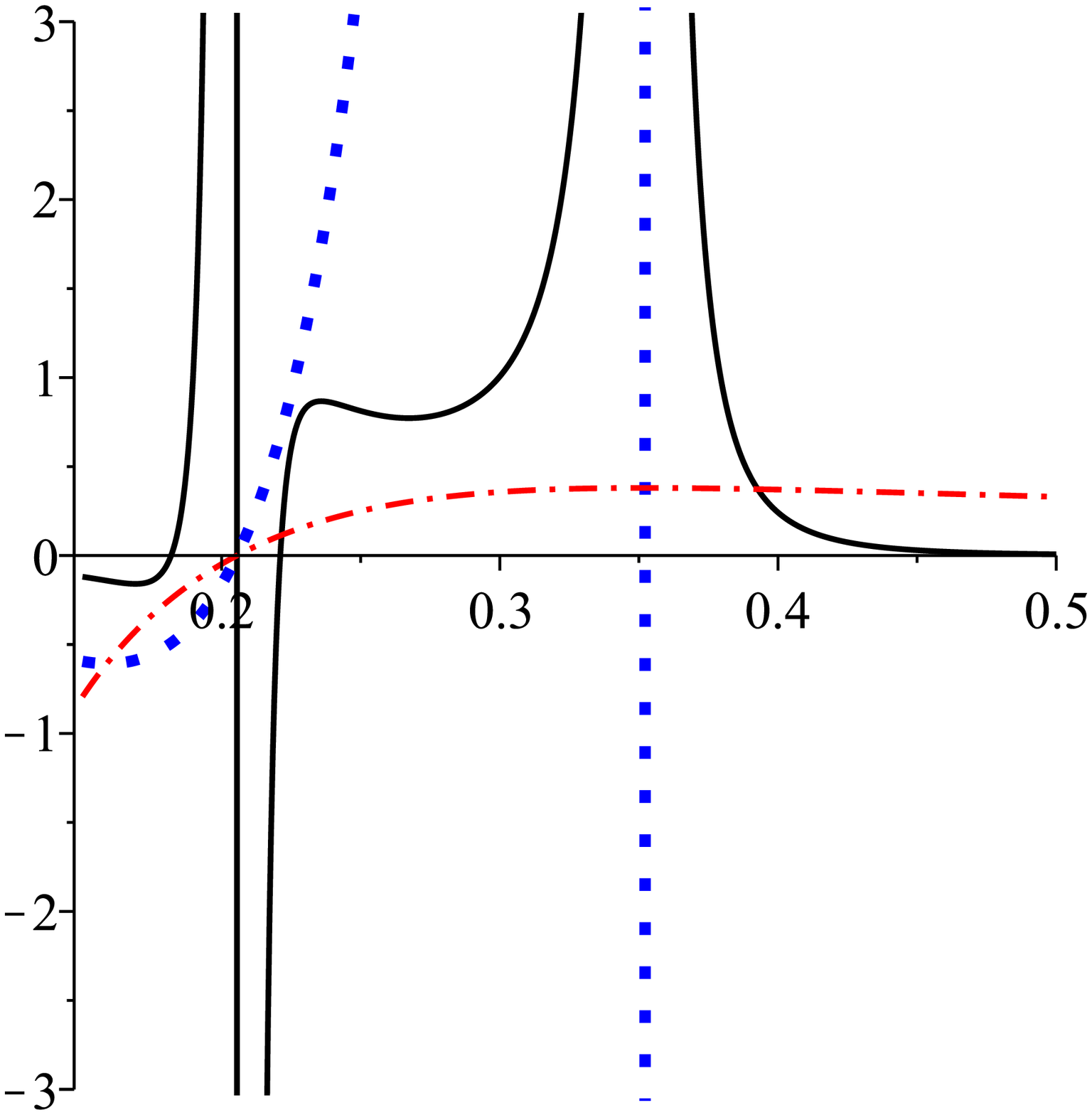} & \epsfxsize=6cm %
\epsffile{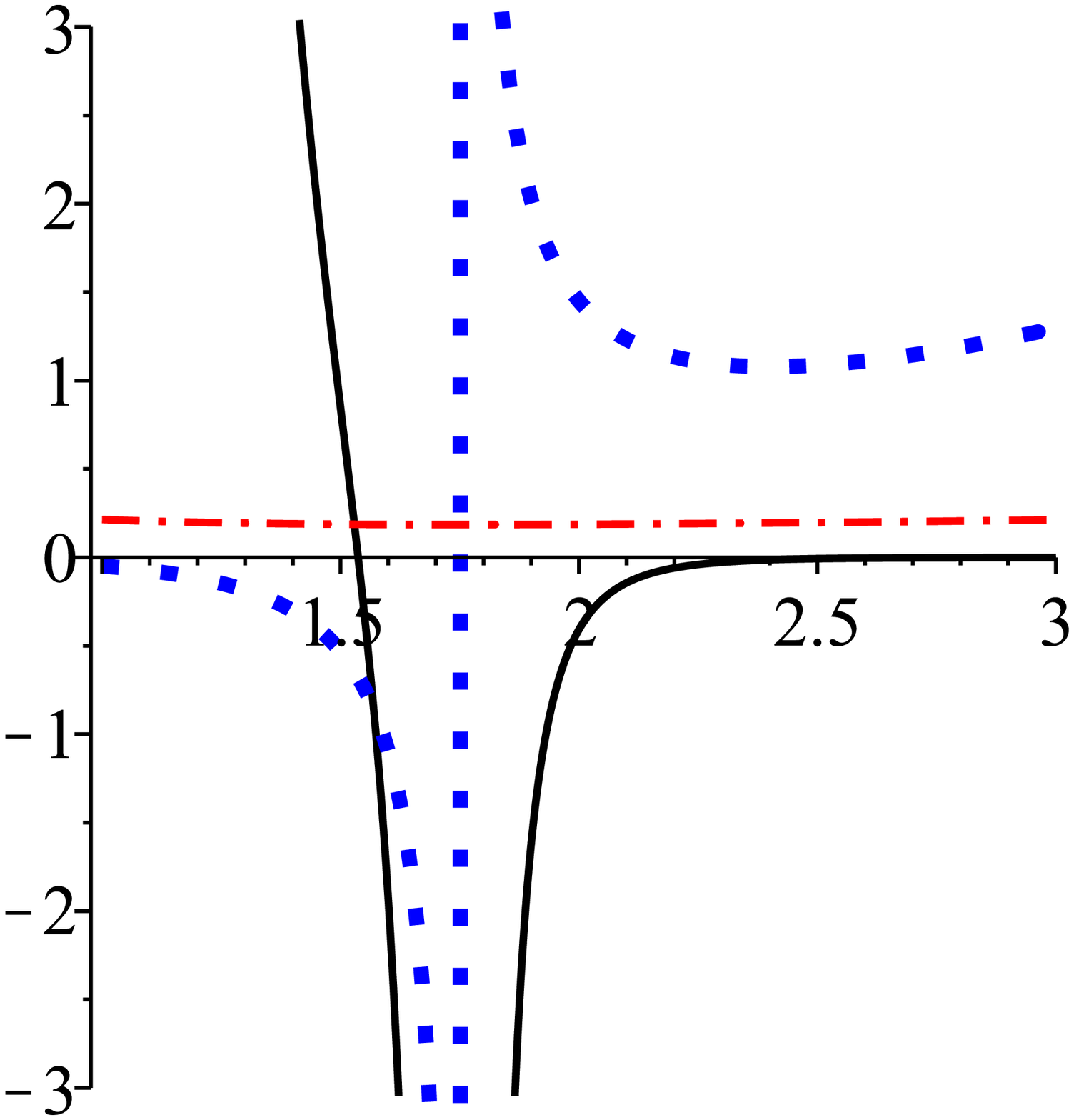}%
\end{array}
$%
\caption{\textbf{HPEM metric:} $\mathcal{R}$ (continuous line),
$C_{Q}$
(dotted line) and {T} (dot dashed) versus $r_{+}$ for $\Lambda =-1$, $n=4$, $%
q=0.1$, $b=1$ and $\protect\omega =10$. $\protect\beta =1$ (up panels) and $%
\protect\beta =5$ (down panels). \emph{"Note: Both panels in the
same line are plotted with the same parameters, but different
regions and scales."}} \label{FigHPEMbeta}
\end{figure}

Very recently, a new metric was proposed by Hendi, et al (HPEM
metric) to solve this problem. This method is applied for various
gravitating systems and it is shown that the root and divergence
points of the heat capacity coincide with the divergence points of
the HPEM Ricci scalar (see Figs.
\ref{FigHPEMbeta}--\ref{FigHPEMn6}, for more details). The
generalized HPEM metric with $n$ extensive variables ($n\geq 2$)
has the following form \cite{HPEM1,HPEM2,HPEM3}
\begin{equation}
ds_{HPEM}^{2}=\frac{SM_{S}}{\left( \Pi _{i=2}^{n}\frac{\partial ^{2}M}{%
\partial \chi _{i}^{2}}\right) ^{3}}\left(
-M_{SS}dS^{2}+\sum_{i=2}^{n}\left( \frac{\partial ^{2}M}{\partial \chi
_{i}^{2}}\right) d\chi _{i}^{2}\right),  \label{HPEM}
\end{equation}
where $\chi_{i}$'s ($\chi_{i}\neq S$) are extensive parameters. It is
notable that HPEM metric is the same as that presented by Quevedo (with the
same $"-,+,+,..."$ signature), but with different conformal factor and
therefore it is expected to enjoy Legendre invariancy. In what follows, we
will investigate the stability and phase transition of the physical BD-BI
black holes in the context of the heat capacity and geometrical
thermodynamics by using HPEM metric.

\begin{center}
\begin{tabular}{c}
\begin{tabular}{cccc}
\hline\hline $n$ \  & $r_{0}$ \  & \ $r_{d_{1}}$ \  & $r_{d_{2}}$
\\ \hline\hline $5$ & $0.2052\ $ & $0.3820$ & $2.5814$ \\ \hline
$6$ \  & $0.2842\ $ & $0.4772\ $ & $3.5605$ \\ \hline $7$\  &
$0.3422\ $ & $0.5390\ $ & $4.7619$ \\ \hline
\end{tabular}
\\
Table I: critical points of BD-BI theory for $q=0.1$, $\Lambda
=-1$, $\omega
=10$, $b=1$ and $\beta =1.5$.%
\end{tabular}

\begin{tabular}{c}
\begin{tabular}{cccc}
\hline\hline $\beta $ & $r_{0}$ \  & \ $r_{d_{1}}$ & $r_{d_{2}}$
\\ \hline\hline
$0.1$\  & $\ 0.0056$ & $0.0112$ & $1.7512$ \\ \hline $1.0$\  & $\
0.0604$ & $0.1251$ & $1.7512$ \\ \hline $1.5$ \  &$\ 0.0902\ $ &
$0.1996$ & $1.7512$ \\ \hline $5.0$ \  & $\ 0.2056$ & $0.3523 $ &
$1.7512$ \\ \hline $100.0$ \  & $\ 0.2400$ & $0.3600 $ & $1.7512$
\\ \hline $200.0$ \  & $\ 0.2400$ & $0.3600 $ & $1.7512$ \\
\hline
\end{tabular}
\\
Table II: critical points of BD-BI theory for $q=0.1$, $\Lambda =-1$, $%
\omega =10$, $b=1$ and $n=4$.%
\end{tabular}

\begin{tabular}{c}
\begin{tabular}{cccc}
\hline\hline $\omega $\  & $r_{0}$ \  & \ $r_{d_{1}}$ \  &
$r_{d_{2}}$ \\ \hline\hline $0.2$\  & $0.0482\ $ & $0.1132$ &
$1.9022$ \\ \hline $2$ \  & $0.0742\ $ & $0.1658\ $ & $1.8038$ \\
\hline $200$ \  & $0.0974\ $ & $0.2128$ & $1.7318$ \\ \hline
\end{tabular}
\\
Table III: critical points of BD-BI theory for $q=0.1$, $\Lambda =-1$, $n=4$%
, $b=1$ and $\beta =1.5$.%
\end{tabular}
\end{center}

\begin{figure}[tbp]
$%
\begin{array}{cc}
\epsfxsize=6cm \epsffile{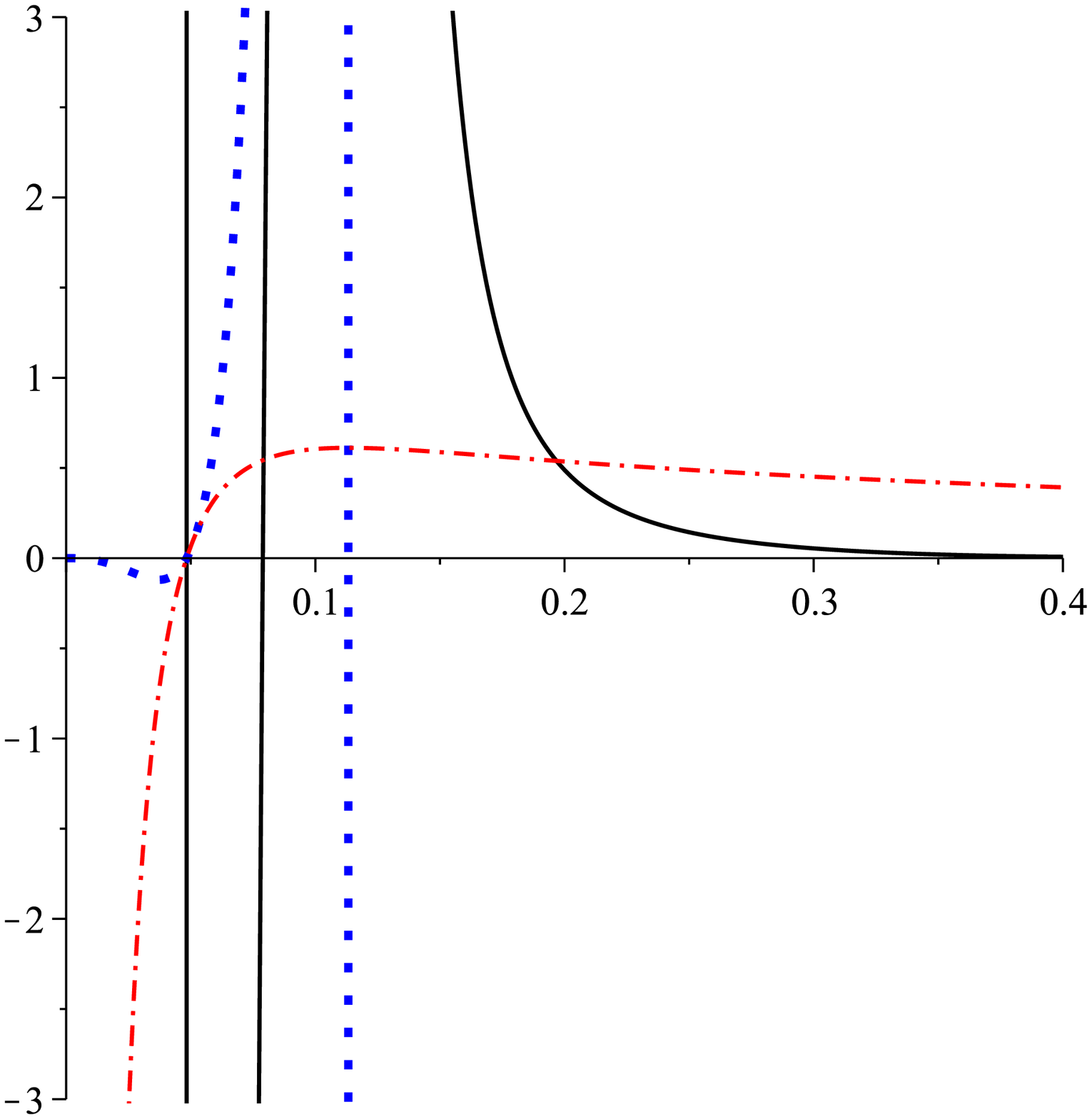} & \epsfxsize=6cm %
\epsffile{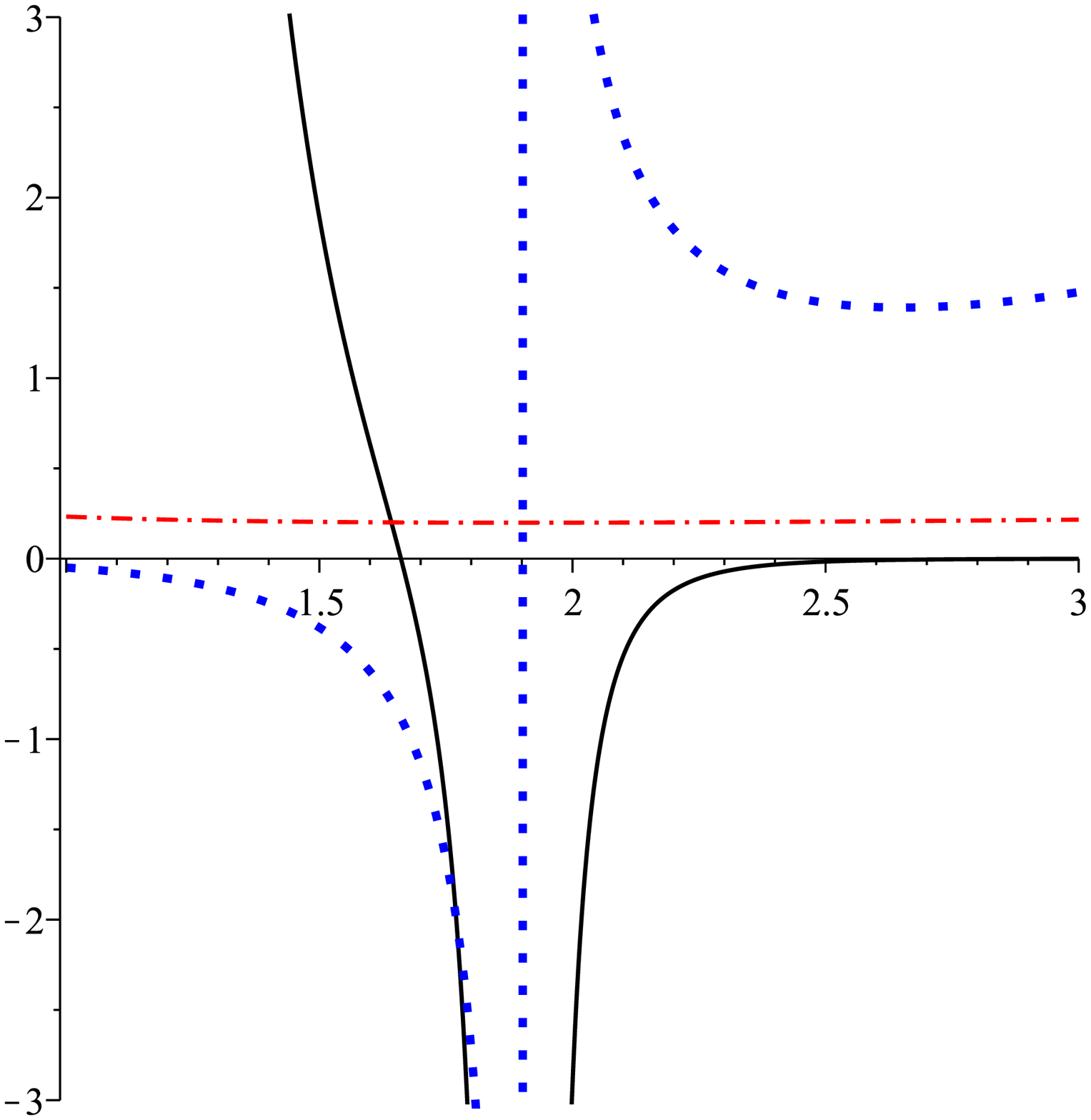} \\
\epsfxsize=6cm \epsffile{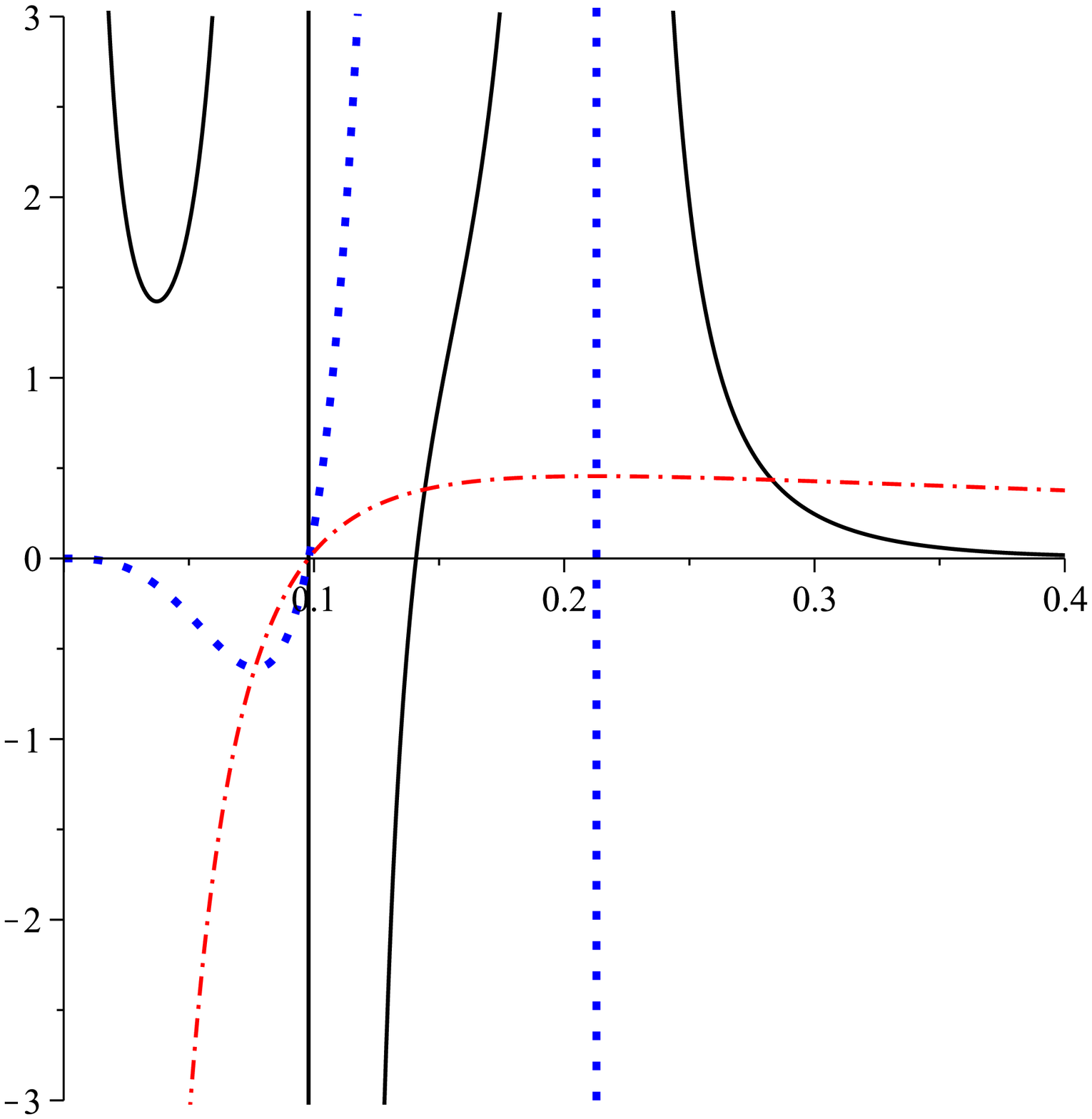} & \epsfxsize=6cm %
\epsffile{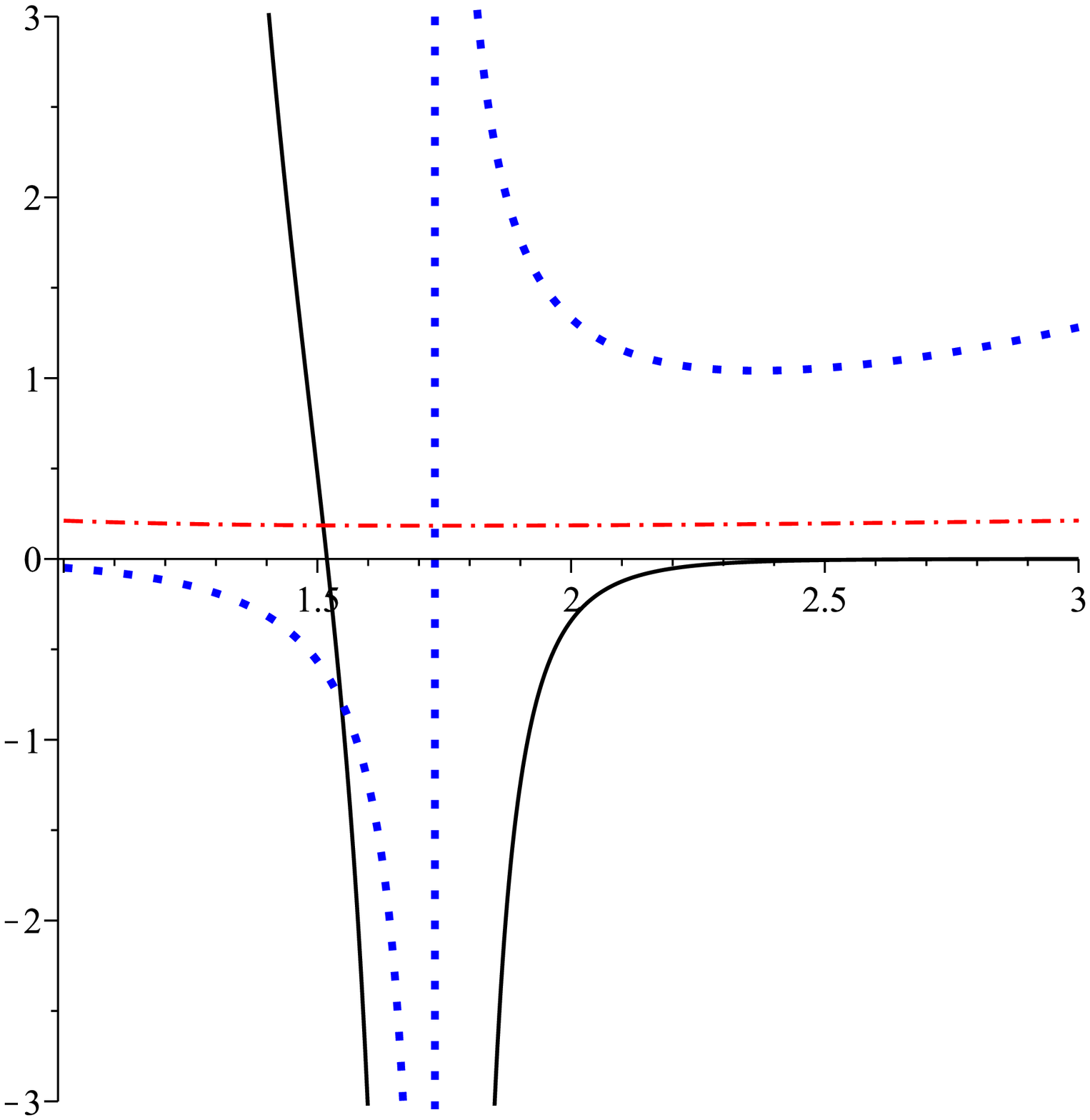}%
\end{array}
$%
\caption{\textbf{HPEM metric:} $\mathcal{R}$ (continuous line),
$C_{Q}$
(dotted line) and {T} (dot dashed) versus $r_{+}$ for $\Lambda =-1$, $n=4$, $%
q=0.1$, $b=1$ and $\protect\beta =1.5$. $\protect\omega =0.2$ (up
panels) and $\protect\omega =200$ (down panels). \emph{"Note: Both
panels in the same line are plotted with the same parameters, but
different regions."}} \label{FigHPEMomega}
\end{figure}

\begin{figure}[tbp]
$%
\begin{array}{cc}
\epsfxsize=7cm \epsffile{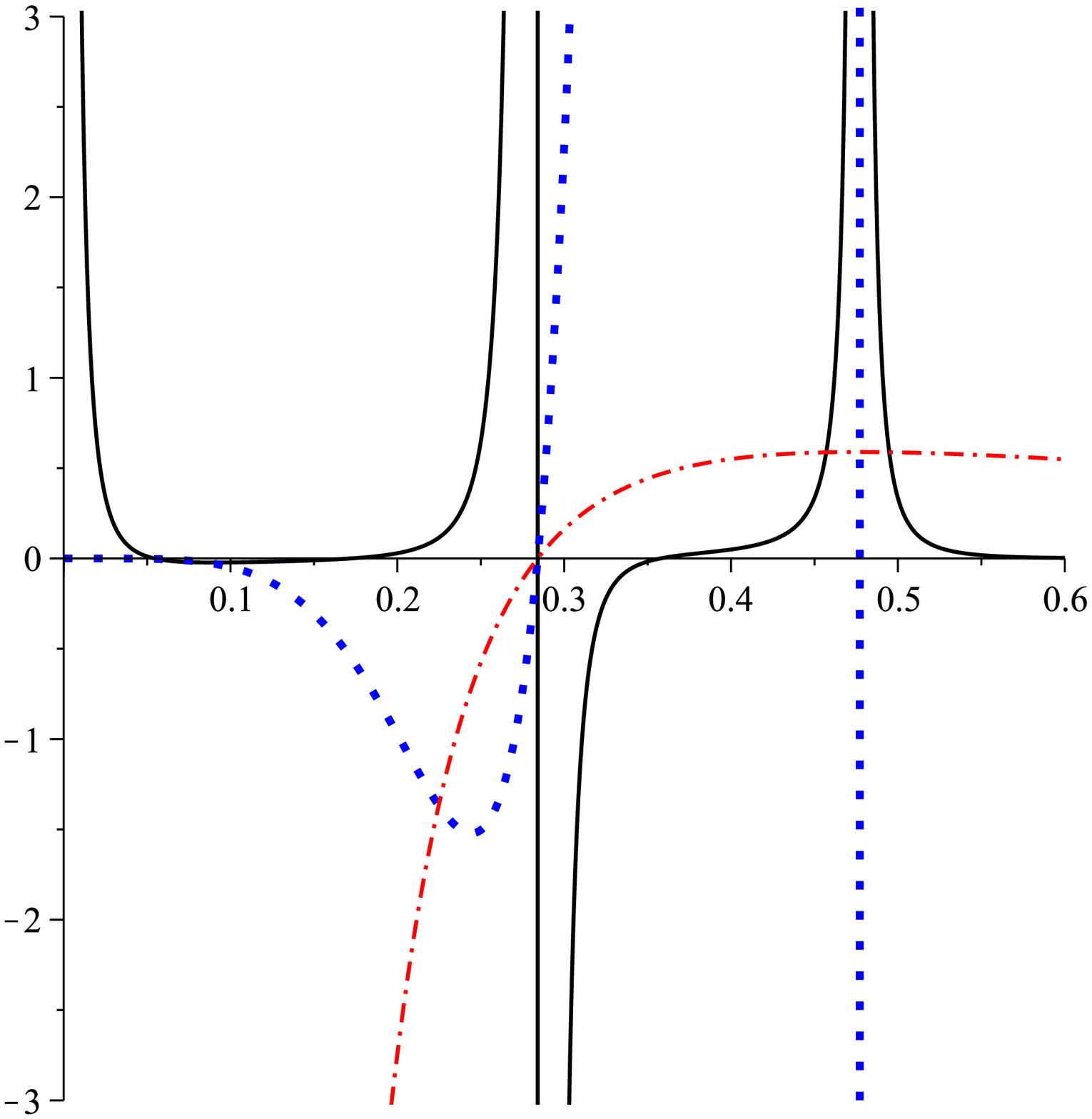} & \epsfxsize=7cm \epsffile{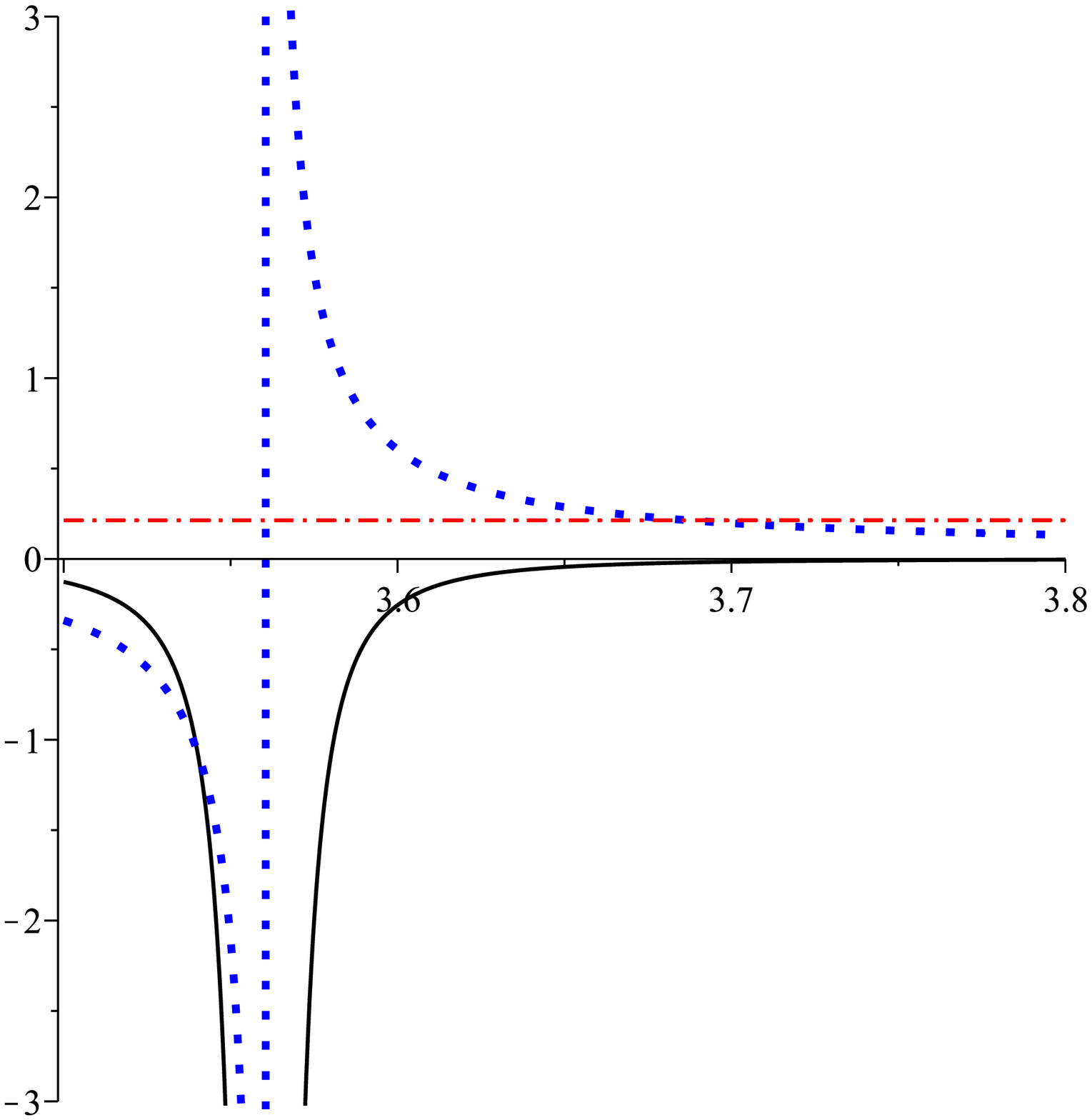}%
\end{array}
$%
\caption{\textbf{HPEM metric:} $\mathcal{R}$ (continuous line),
$C_{Q}$
(dotted line) and {T} (dot dashed) versus $r_{+}$ for $\Lambda =-1$, $n=6$, $%
q=0.1$, $b=1$, $\protect\omega =10$ and $\protect\beta =1.5$.
\emph{"Note: Both panels are plotted with the same parameters, but
different regions and scales."}} \label{FigHPEMn6}
\end{figure}

\begin{figure}[tbp]
$%
\begin{array}{ccc}
\epsfxsize=5.6cm \epsffile{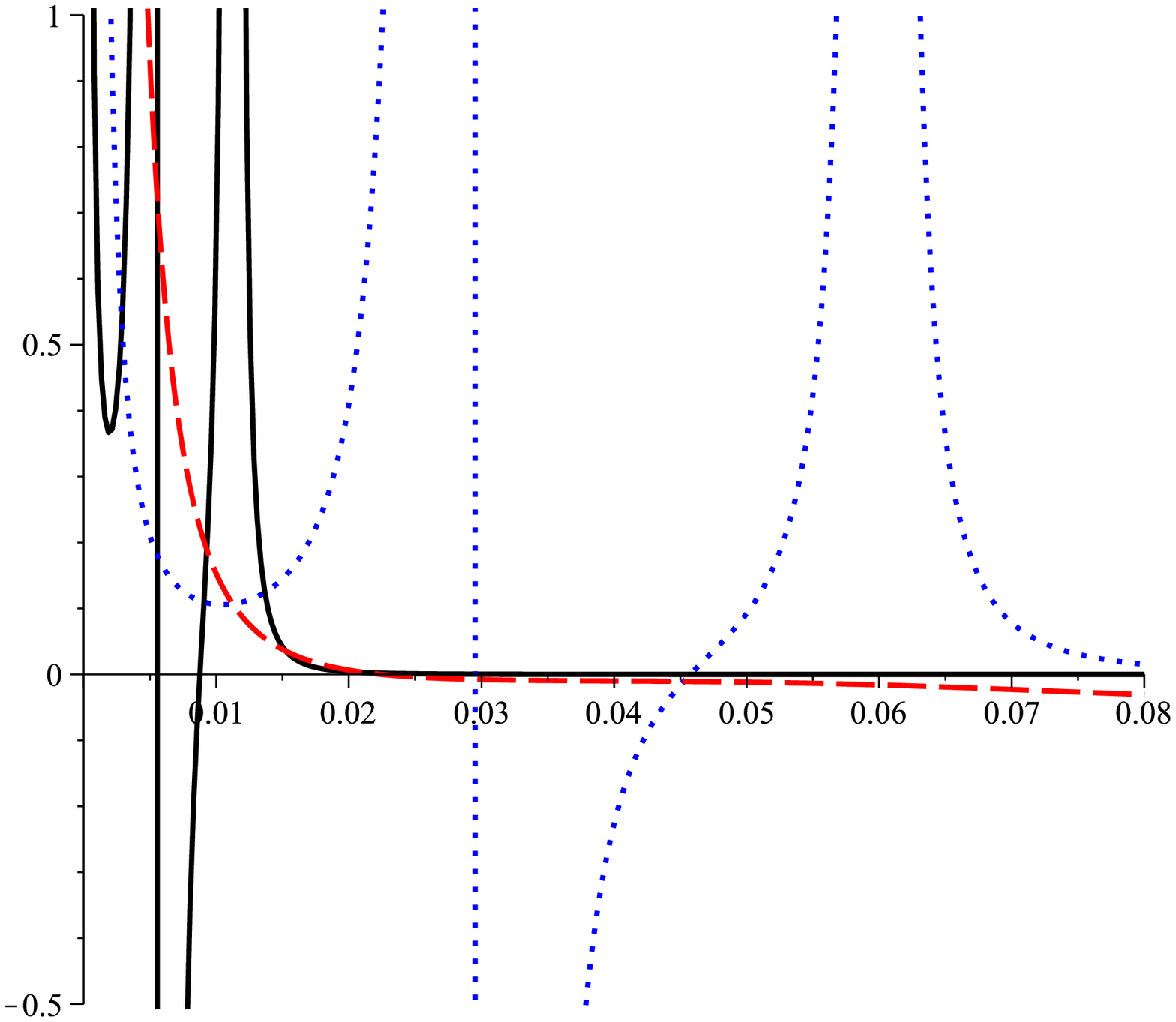} & \epsfxsize=5.6cm \epsffile{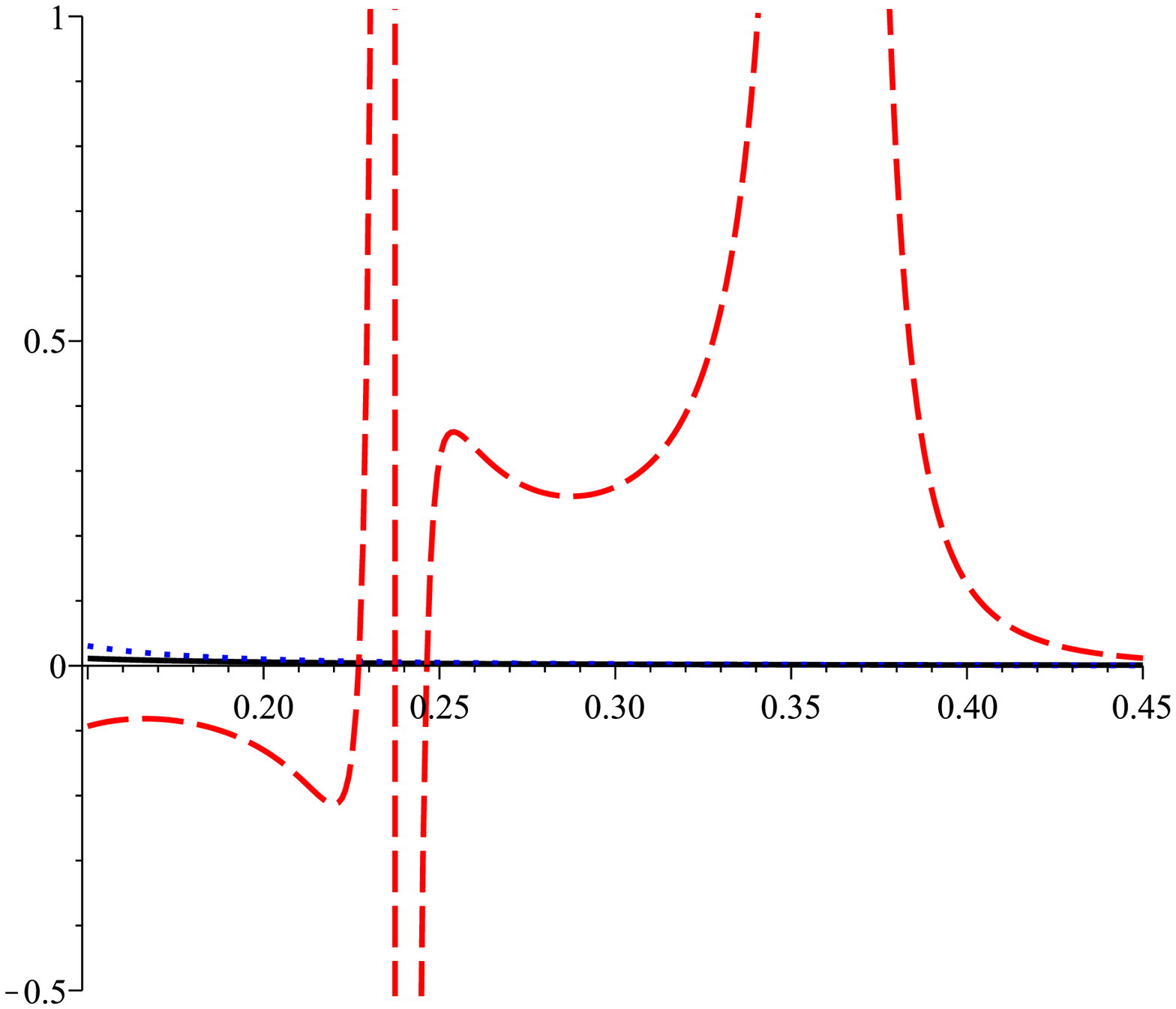} & \epsfxsize=5.6cm \epsffile{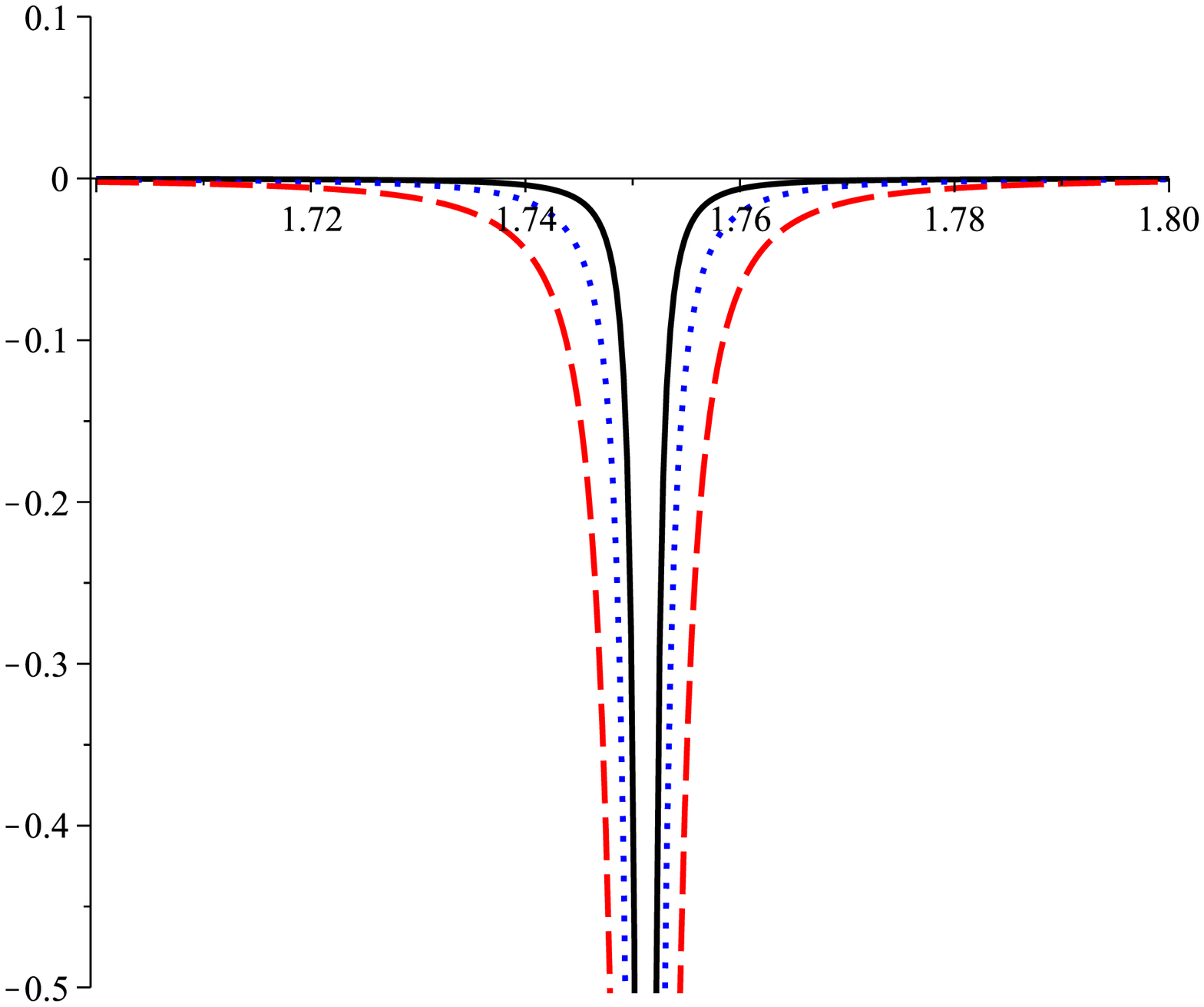} %
\end{array}
$%
\caption{\textbf{HPEM metric:} $\mathcal{R}$ versus $r_{+}$ for $\Lambda =-1$, $n=4$, $%
q=0.1$, $b=1$, $\protect\omega =10$ and $\protect\beta =0.1$
(continuous line), $\protect\beta =0.5$ (dotted line) and Maxwell
(dot dashed). \emph{"Note: All three panels are plotted with the
same parameters, but different regions and scales."}}
\label{FigHPEMomega}
\end{figure}

As the first significant point which must be taken into deep
consideration, one should regard the sign of the temperature. The
positivity of the temperature denotes a physical black hole;
Whereas the negativity of $T$ represents a non-physical system.
The temperature behavior has been shown in figures, too. As we can
see, there is a lower bound for the horizon radius ($r_{0}$), in
which for $r_{+}< r_{0}$, we encounter with a non-physical black
hole, owing to negative sign of temperature. In contrast, in the
case of $r_{+}> r_{0}$, we confront a physical system due to the
positivity of the temperature. In other words, the horizon radius
of physical black holes are located in this region.

Figure \ref{FigHPEMbeta} shows that for the special values of the
electric charge, nonlinearity parameter and BD-coupling
coefficient, we can obtain three characteristic points. One of
them refers to the root of heat capacity (or temperature) which is
known as $r_{0}$ and others are related to the divergence points
of heat capacity which are denoted as $r_{d_{1}}$and $r_{d_{2}}$
($r_{d_{1}} < r_{d_{2}}$). We also find that all divergence points
of the Ricci scalar (for HPEM metric) are coincide with these
three points. Here, we use some tables to study the influences of
different parameters (dimensions, nonlinearity parameter and
BD-coupling coefficient) on the mentioned characteristic points.

These tables provide information regarding the lower bound of
horizon radius, two points of phase transition (for the case of
BD-BI) and their dependencies to the variation of dimensions,
nonlinearity parameter and coupling coefficient. Regarding the
tables and Figs. \ref{FigHPEMbeta}--\ref{FigHPEMomega}, it is
evident that one root and two divergence points for the heat
capacity are almost observed. It is worthwhile to mention that,
the region of $r_{0}<r_{+}<r_{d_1}$, (positive sign of heat
capacity) shows the stability of the system. In contrast, one can
find that for the region of $r_{d_1}<r_{+}<r_{d_2}$, the heat
capacity has negative sign which indicates instability. In
addition, at region $r_{+}>r_{d_2}$, the system is in the stable
state due to the positive sign of heat capacity (see Figs. 4-6 for
more details). According to table I, one can conclude that the
lower bound radius and two divergence points are increasing
functions of the dimensions. Also, according to table II, the
lower bound of horizon radius and the first divergence point
($r_{d_1}$) will increase by increasing $\beta $ (the nonlinearity
parameter), whereas the second point of divergency remains steady
over this change. Considering figures and table II, it is obvious
that by increasing $\beta$, root and the first divergence point of
heat capacity will increase up to a point and then any increment
in this parameter would have negligible effect on these values. To
put in other words, it can be interpreted that in large $\beta $,
we will face the Brans-Dicke-Maxwell behavior \cite{BD-Max}. For
large $\beta $, the obtained values for lower bound horizon radius
and divergence point are the same as the obtained values for the
Brans-Dicke-Maxwell case \cite{BD-Max}. It is notable that the
unstable region (between two divergencies, where the heat capacity
is negative) is larger in small $\beta $ than the large one
(Brans-Dicke-Maxwell case) as it would be expected, which is due
to the nature of nonlinearity that would cause the instability of
system to increase. Meanwhile, $r_{0}$ and $r_{d_1}$ have
ascending functions and $r_{d_2}$ will be declined by increasing
$\omega $ (see table III). Generally, from what has been discussed
above, dimensionality $n$ and BD-coupling coefficient $\omega $
are playing the main role in changes of the location of larger
divergence point.

\section{Conclusion}

In this paper, the main goal was studying thermodynamical behavior
of the BD-BI and Einstein-BI-dilaton black hole solutions. Since
both of these solutions had very similar thermodynamical behavior
in the context of geometrical thermodynamics, we have just
considered the BD-BI ones. We have investigated the stability and
phase transition in the canonical ensemble through the use of heat
capacity. We have found that for having a physical black hole
(positive temperature), there should be a restriction on the value
of the horizon radius, which lead to a physical limitation point.
This point was a border between non-physical and physical black
hole horizon radius. Moreover, investigating the phase transition
of the black holes exhibited that there exist second order phase
transition points. In other words, the heat capacity had one real
positive root and two divergence points. It was shown that these
points (the root and divergence points of heat capacity) were
affected by variation of the BI-parameter, BD coupling constant
and dimensions. From presented tables and figures, we have found
that the effect of dimensions on the larger divergence point was
more than other factors and in contrast, the BI-parameter had no
sensible effect on this value. The effect of BD-coupling constant
on these three points was so small in a way that by applying a
dramatic change in this constant, we observed a small change in
the value of such characteristic points.

It was illustrated that in the context of thermal stability there
exist four regions, specified by the root and two divergence
points of the heat capacity. The root of heat capacity was
referred as the lower bound of horizon radius that separated the
non-physical black holes from the physical ones. Between the two
divergencies, we encountered an unstable state and after the
second divergence point black hole obtained a stable state. It is
notable that for small $\beta$, because of the nonlinearity
effect, the unstable region is larger than the Maxwell case (large
$\beta$) \cite{BD-Max}.

Eventually, we employed the geometrical thermodynamic method to
study the phase transition. We have shown that Weinhold, Ruppeiner
and Quevedo metrics failed to provide a consistent result with the
heat capacity's result. In other words, their thermodynamical
Ricci scalar's divergencies did not match with the root and
divergencies of the heat capacity, exactly. In some of these
methods we encountered extra divergency which did not coincide
with any of the phase transition points.

At last, using the HPEM metric, we achieved desirable results. It
was shown that all the divergence points of the Ricci scalar of
the mentioned metric covered the divergencies and root of the heat
capacity. It is worth mentioning that the behavior of the
curvature scalar was different near its divergence points. In
other words, the divergence points of the Ricci scalar related to
root of the heat capacity could be distinguished from the
divergencies related to phase transition points based on the
curvature scalar behavior.

Regarding the used method of this paper, it is interesting to
extend obtained results to an energy dependent spacetime and
discuss the role of gravity's rainbow
\cite{raibow1,rainbow2,rainbow3,rainbow4}. We leave this issue for
future work.

\begin{center}
\textbf{Competing Interests}
\end{center}
The authors declare that there is no conflict of interests
regarding the publication of this paper.

\begin{acknowledgements}
We would like to thank the anonymous referee for his valuable
comments. We also acknowledge M. Momennia and S. Panahiyan for
reading the manuscript. We wish to thank Shiraz University
Research Council. This work has been supported financially by the
Research Institute for Astronomy and Astrophysics of Maragha
(RIAAM), Iran.
\end{acknowledgements}

\end{document}